\documentclass[iop]{emulateapj}
  
\usepackage{apjfonts}
\usepackage{amsmath}
\usepackage{epsf}
\usepackage{color}
\shorttitle{CANDELS: Galaxy Colors from $z =$ 4 to 8}
\shortauthors{Finkelstein et al.}

\newcommand{\sol}{$_{\odot}$}
\newcommand{\lya}{Ly$\alpha$}

\def\arcs{\hbox{$^{\prime\prime}$}}

\begin{document}
\slugcomment{Accepted to the Astrophysical Journal}
\title{CANDELS: The Evolution of Galaxy Rest-Frame Ultraviolet Colors from $\MakeLowercase{z}$ = 8 to 4}

\author{Steven   L.    Finkelstein\altaffilmark{1,2,a}, Casey Papovich\altaffilmark{2}, Brett Salmon\altaffilmark{2}, Kristian Finlator\altaffilmark{3}, Mark Dickinson\altaffilmark{4}, Henry C. Ferguson\altaffilmark{5}, Mauro Giavalisco\altaffilmark{6}, Anton M. Koekemoer\altaffilmark{5}, Naveen A. Reddy\altaffilmark{7}, Robert Bassett\altaffilmark{2}, Christopher J. Conselice\altaffilmark{8}, James S. Dunlop\altaffilmark{9}, S. M. Faber\altaffilmark{10}, Norman A. Grogin\altaffilmark{5}, Nimish P. Hathi\altaffilmark{11}, Dale D. Kocevski\altaffilmark{10}, Kamson Lai\altaffilmark{10}, Kyoung-Soo Lee\altaffilmark{12}, Ross J. McLure\altaffilmark{9}, Bahram Mobasher\altaffilmark{7} \& Jeffrey A. Newman\altaffilmark{13}}

\altaffiltext{1}{Department of Astronomy, The University of Texas at Austin, Austin, TX 78712}
\altaffiltext{2}{George P. and Cynthia Woods Mitchell Institute for Fundamental Physics and Astronomy, Department of Physics and Astronomy, Texas A\&M University, College Station, TX 77843}
\altaffiltext{3}{Physics Department, University of California, Santa Barbara, CA 93106}
\altaffiltext{4}{National Optical Astronomy Observatory, Tucson, AZ 85719}
\altaffiltext{5}{Space Telescope Science Institute, Baltimore, MD 21218}
\altaffiltext{6}{Department of Astronomy, University of Massachusetts, Amherst, MA 01003}
\altaffiltext{7}{Department of Physics and Astronomy, University of California, Riverside, CA 92521}
\altaffiltext{8}{University of Nottingham, School of Physics \& Astronomy, Nottingham, UK}
\altaffiltext{9}{Institute for Astronomy, University of Edinburgh, Royal Observatory, Edinburgh, UK}
\altaffiltext{10}{University of California Observatories/Lick Observatory, University of California, Santa Cruz, CA, 95064}
\altaffiltext{11}{Carnegie Observatories, Pasadena, CA 91101}
\altaffiltext{12}{Yale Center for Astronomy and Astrophysics, Departments of Physics and Astronomy, Yale University, New Haven, CT 06520}
\altaffiltext{13}{Department of Physics and Astronomy, University of Pittsburgh, Pittsburgh, PA 15260}
\altaffiltext{a}{Hubble Fellow, stevenf@astro.as.utexas.edu}

\begin{abstract}
We study the evolution of galaxy rest-frame ultraviolet (UV) colors in the epoch 4 $\lesssim z \lesssim$ 8.  We use new wide-field near-infrared data in the Great Observatories Origins Deep Survey -- South field from the Cosmic Assembly Near-infrared Deep Extragalactic Legacy Survey, Hubble Ultra Deep Field 2009 and Early Release Science programs to select galaxies via photometric redshift measurements.  Our sample consists of 2812 candidate galaxies at $z \gtrsim$ 3.5, including 113 at $z \simeq$ 7-- 8.  We fit the observed spectral energy distribution (SED) to a suite of synthetic stellar population models, and measure the value of the UV spectral slope ($\beta$) from the best-fit model spectrum.  We run simulations to show that this measurement technique results in a smaller scatter on $\beta$ than other methods, as well as a reduced number of galaxies with catastrophically incorrect $\beta$ measurements (i.e., $\Delta\beta >$ 1).  We find that the median value of $\beta$ evolves significantly from $-$1.82$^{+0.00}_{-0.04}$ at $z =$ 4, to $-$2.37$^{+0.26}_{-0.06}$ at $z =$ 7.  Additionally, we find that faint galaxies at $z =$ 7 have $\beta$ = $-$2.68$^{+0.39}_{-0.24}$ ($\sim -$2.4 after correcting for observational bias); this is redder than previous claims in the literature, and does not require ``exotic'' stellar populations (e.g., very-low metallicities or top-heavy initial mass functions) to explain their colors.  This evolution can be explained by an increase in dust extinction, from low amounts at $z =$ 7, to A$_{\mathrm{V}} \sim$ 0.5 mag at $z =$ 4.  The timescale for this increase is consistent with low-mass AGB stars forming the bulk of the dust.  We find no significant ($<$ 2$\sigma$) correlation between $\beta$ and M$_{UV}$ when measuring M$_{UV}$ at a consistent rest-frame wavelength of 1500 \AA.  This is particularly true at bright magnitudes, though our results do show evidence for a weak correlation at faint magnitudes when galaxies in the HUDF are considered separately, hinting that dynamic range in sample luminosities may play a role.  We do find a strong correlation between $\beta$ and the stellar mass at all redshifts, in that more massive galaxies exhibit redder colors.  The most massive galaxies in our sample have similarly red colors at each redshift, implying that dust can build up quickly in massive galaxies, and that feedback is likely removing dust from low-mass galaxies at $z \geq$ 7.  Thus the stellar-mass -- metallicity relation, previously observed up to $z \sim$ 3, may extend out to $z =$ 7 -- 8.
\end{abstract}

\section{Introduction}
Recent advances in technology, including the installation of the near-infrared (NIR) channel of the Wide Field Camera 3 (WFC3) on board the {\it Hubble Space Telescope} ({\it HST}), have revolutionized our ability to study the early universe.  Initial WFC3 programs, including the deepest NIR image ever taken in the Hubble Ultra Deep Field (HUDF; ID 11563, PI Illingworth), allowed the discovery of the first robust samples of galaxies at $z >$ 7 \citep{bouwens10a, oesch10, mclure10, bunker10, finkelstein10a, yan10}.  In addition to the exciting discovery that large samples of galaxies exist at such early times, these data allowed us to study the physical properties of these galaxies, at a time only $\sim$ 500--800 Myr removed from the Big Bang.

At $z \sim$ 7 -- 8, these WFC3 NIR data are probing the rest-frame ultraviolet (UV) light from these galaxies.  While the {\it Spitzer Space Telescope} can observe the rest-frame optical light, its sensitivity is not at the level to detect the typical, sub-L$^{\ast}$ galaxies that make up most of the published samples (although stacking has resulted in some detections \citep{labbe10}, one must be cautious at these redshifts as high equivalent width H$\alpha$ and [O\,{\sc iii}] emission can dominate fluxes in the {\it Spitzer} bands \citep[e.g.][]{finkelstein09a, schaerer10, shim11}).  However, the rest-frame ultraviolet alone is an important diagnostic of galaxy physical properties. 

\citet{calzetti94} parameterized the UV color via the UV spectral slope, $\beta$, defined as f$_{\lambda}$ $\propto$ $\lambda^{\beta}$, which was designed to be measured from spectra in specified wavelength windows.  Subsequent work has found that $\beta$ is an excellent tracer of the dust extinction in galaxies, as it is well correlated with dust emission in the far-infrared (FIR) (defined as the ratio of FIR-to-UV flux), at both low redshift \citep{meurer97, meurer99}, and $z \sim$ 2 \citep[e.g.,][]{seibert02, reddy12}.

Studying just the rest-frame UV color at high redshift can thus provide powerful information on the presence of dust in galaxies, which in turn provides insight into the chemical enrichment of the universe.  However, at such high redshifts continuum spectroscopy is beyond our grasp (though the upcoming \textit{James Webb Space Telescope}  should provide this capability), thus $\beta$ is traditionally estimated via a single UV color \citep[e.g.][]{meurer97, hathi08, overzier08, bouwens10b, finkelstein10a, dunlop12}.

Using a single color, \citet{bouwens09} measured $\beta$ for a sample of $\sim$ 1400 galaxies from 2 $< z <$ 6.  They found strong evolution, in that $\beta$ changed from $\sim -$1.5 at $z \sim$ 2 to $\sim -$2 at $z \sim$ 6, and found also that lower luminosity galaxies appeared to be bluer at a given redshift.  This implies that the dust content in galaxies is decreasing towards earlier epochs.  This leads to the natural question of whether galaxies get even bluer at higher redshifts.  \citet{bouwens10b} and \citet{finkelstein10a} measured $\beta$ for galaxies at $z \sim$ 7, finding that the fainter (m$_{F160W} \gtrsim$ 28.3) galaxies exhibit $\beta$ $\sim -$3 in both cases.  \citet{bouwens10b} use a measured small uncertainty ($\Delta\beta \sim$ 0.2) to investigate the possibility of the presence of extremely metal-poor (or perhaps metal-free) stars, while \citet{finkelstein10a} argue for a larger uncertainty ($\Delta\beta \sim$ 0.5), and concluded that the colors of these galaxies are consistent with stellar populations observed locally.  Recently, \citet{mclure11} and \citet{dunlop12} have analyzed a sample of galaxies at $z \sim$ 7, and found a variance-weighted mean value of $\beta \sim -$2.0.  However, we note that this study used more rigorous selection criteria that excluded many of the fainter galaxies, which may explain the apparent redder value of the UV spectral slope.  In any case, the differing results from previous studies highlight that more data are clearly needed to understand how galaxy colors are evolving out to the highest redshifts.

In this paper, we use recently obtained wide field WFC3 data in conjunction with the deep HUDF data to compile a much larger sample of galaxies at $4 < z < 8$ to make a more robust determination of the evolution of $\beta$.  In \S 2 we describe the data used and catalog construction, as well as our sample selection techniques.  In \S 3 we discuss our new method for measuring the UV spectral slope $\beta$, as well as the simulations we used to test our method.  In \S 4 we discuss the evolution of $\beta$ with redshift, while in \S 5 we examine whether $\beta$ is correlated with the UV luminosity or the stellar mass, with the implications discussed in \S 6.  Our conclusions are presented in \S 7.  All magnitudes are in the AB system, and throughout we assume a concordance cosmology with H$_\mathrm{o}$ = 70 km s$^{-1}$ Mpc$^{-1}$, $\Omega_{m}$ = 0.3 and $\Omega_{\Lambda}$ = 0.7.

\section{Observations}

\subsection{Data}

We select our sample of galaxies from a number of datasets.  First, we use recently obtained data from the Cosmic Assembly Near-Infrared Deep Extragalactic Survey\footnote[1]{http://candels.ucolick.org} \citep[CANDELS; Co-PIs Faber \& Ferguson;][]{grogin11, koekemoer11}, which has obtained deep WFC3/IR imaging of the central $\sim$50\% of GOODS-S (hereafter referred to as GOODS-S DEEP), and shallower WFC3/IR imaging of the southern $\sim$25\% of GOODS-S (hereafter referred to as GOODS-S WIDE).  We also use the recent {\it HST}/WFC3 imaging data of the HUDF main field.  We now use all of the images of this field from the completed HUDF09 program, which are $\sim$ 0.2--0.3 mag deeper in the F105W, F125W and F160W bands in contrast to the initial papers using the HUDF09 data  \citep[e.g.,][]{oesch10,bouwens10a}.  Finally, we use the data from the WFC3 Early Release Science \citep[ERS;][]{windhorst11} program, which imaged the northern $\sim$25\% of the Great Observatories Origins Deep Survey \citep[GOODS][]{dickinson03, giavalisco04} Chandra Deep Field -- S (hereafter, GOODS-S) with WFC3.  Figure 15 of \citet{koekemoer11} highlights the relative positions of these fields.  Each of these four fields contains imaging in the F105W, F125W and F160W bands (hereafter Y$_{105}$, J$_{125}$ and H$_{160}$, respectively), except for the ERS, which uses the narrower F098M (hereafter Y$_{098}$) in place of F105W.  All of these data have been reprocessed in the same way; details can be found in \citet{koekemoer11}.  Throughout the paper, the term ``Y-band'' will denote either Y$_{098}$ or Y$_{105}$, depending on the field.

We use archival {\it HST} optical data obtained with the Advanced Camera for Surveys in the F435W, F606W, F775W and F850LP filters (hereafter referred to as B$_{435}$, V$_{606}$, i$_{775}$ and z$_{850}$) from \citet{beckwith06} for the HUDF, as well as the GOODS ACS v2 imaging for the DEEP, WIDE and ERS fields.  All of the data has been drizzled to a pixel scale of 0.06\arcs.  The survey areas and imaging depths probed by these fields are summarized in Table 1.

\subsection{Object Photometry}

The size of the point-spread function (PSF) varies across our dataset due to the long wavelength baseline covered ($\sim$ 0.4 - 1.8 $\mu$m).  Thus, prior to photometering our imaging, we match the PSF of each image to the H$_{160}$-band, which has the largest PSF full-width at half-maximum (FWHM).  First, we construct empirical PSFs by stacking stars identified in each image, subsampling to ensure accurate alignment, and then sampling back up to the original pixel scale of 0.06\arcs.  We use the DECONV$\_$TOOL routine in IDL \citep{varosi93} to create a convolution kernel from the PSF of each image and the H$_{160}$-band PSF.  This kernel is then convolved with each image.  To ensure that the PSF of the resulting images matches that of the H$_{160}$-band, we measure the curves-of-growth of stars in each image.  We iterate this procedure until the resulting curves-of-growth are within 1\% of the H$_{160}$-band at aperture diameters larger than 0.4\arcs\ (similar to the expected sizes of our objects) for all bands\footnote[2]{Except for B$_{435}$, which was only able to be matched within $<$ 8\% of the H$_{160}$-band PSF at an aperture diameter of 0.4\arcs\ (the HUDF, GOODS DEEP, WIDE and ERS fields had B$_{435}$ curves-of-growth matched to within 3\%, 4\%, 6\% and 8\%, respectively).}.

\begin{deluxetable*}{cccccccccc}
\tabletypesize{\small}
\tablecaption{Observations Summary}
\tablewidth{0.8\textwidth}
\tablehead{
\colhead{Field} & \colhead{Area} & \colhead{B$_{435}$} & \colhead{V$_{606}$} & \colhead{i$_{775}$}  & \colhead{z$_{850}$} & \colhead{$Y_{098}$} & \colhead{$Y_{105}$} & \colhead{$J_{125}$} & \colhead{$H_{160}$}\\
\colhead{$ $} & \colhead{(arcmin$^{2}$)} & \colhead{(mag)} & \colhead{(mag)} & \colhead{(mag)} & \colhead{(mag)} & \colhead{(mag)} & \colhead{(mag)} & \colhead{(mag)} & \colhead{(mag)}\\
}
\startdata
HUDF&5.0&29.5&29.9&29.7&29.0&---&29.2&29.5&29.5\\
GOODS-S DEEP$^{\ddagger}$&85.6&28.1&28.3&27.7&27.5&---&28.0&27.7&27.5\\
GOODS-S WIDE&40.1&28.1&28.3&27.7&27.5&---&27.4&27.6&27.3\\
GOODS-S ERS&44.7&28.1&28.3&27.7&27.5&27.6&---&28.0&27.7
\enddata
\tablecomments{The area is measured from the image area used to
  detect objects, which excludes noisy regions on the edges.   The
  remaining columns are 5$\sigma$ limiting magnitudes measured in
  a 0.4\arcs-diameter apertures on non-PSF matched images.  $^{\ddagger}$ The  CANDELS GOODS-S DEEP  field data used
  in  this analysis represent  the first  four (out  of 10)  epochs of
  J$_{125}$ and H$_{160}$ data.  Additionally, although it is to full
  depth, the CANDELS-DEEP Y$_{105}$  band data used covers only  $\sim$ 33\% of the
  DEEP  region (the  remainder will  be imaged  in a  second  epoch in
  December 2011).  We note that we selected objects over the full DEEP
  region; thus $\sim$ 67\% of our CANDELS-DEEP objects do not yet have
  Y$_{105}$ photometry.  }
\end{deluxetable*}

We extract photometry for sources in these images by first creating a detection image for each of the four fields via an inverse-variance weighted sum of the J$_{125}$ and H$_{160}$ images.  We then use a modified version of the Source Extractor (hereafter SExtractor) software \citep[v2.8.6;][]{bertin96} to perform the photometry, which adds a buffer between the source and the local background cell, and removes spurious sources associated with the distant wings of bright objects.  We run the modified SExtractor in two-image mode, using the J+H image as the detection image, and then successively using the B$_{435}$, V$_{606}$, i$_{775}$, z$_{850}$, Y$_{098/105}$, J$_{125}$ and H$_{160}$ images as the measurement image.  We used the MAP$\_$RMS weighting option, and we provided accurate rms maps for each image.  We measured photometry in small elliptical apertures, using the PHOT\_AUTOPARAMS values of 1.2, 1.7 (similar to that done in Finkelstein et al.\ 2010), as well as in a series of circular apertures.  To test the photometric uncertainties returned by SExtractor, we measured the noise in the unsmoothed images in randomly placed 0.4\arcs-diameter apertures.  We found that this noise was consistent with the photometric errors in the same size apertures returned by SExtractor, highlighting that the SExtractor-derived uncertainties are accurate.
 
To select apertures appropriate for our science goals, we ran simulations in the HUDF data inserting mock galaxies into the images and measuring their photometry in the same manner as described above (these simulations are described in more detail in \S 3.2).  We then compare the measured colors of these mock galaxies to the input color in both the small elliptical apertures, as well as in circular apertures with diameters of 0.4\arcs\ and 0.6\arcs, as a function of magnitude.  The scatter between the input and recovered colors for both the small elliptical and the 0.4\arcs-diameter circular apertures are nearly identical over the entire magnitude range, while the 0.6\arcs-diameter apertures begin to have significantly worse scatter in the color at $m >$ 28 (with a scatter $\sim$ 0.05 higher than the other apertures at H$_{160} =$ 29 mag).  Comparing the elliptical and the 0.4\arcs-diameter circular apertures, the latter yield slightly better (i.e., 0.01 mag) results at relatively bright magnitudes (m$_{H}$ $\sim$ 26--27), while the small elliptical apertures yield slightly better (i.e., 0.02 mag) results at fainter magnitudes.  We then turned to our object catalog, and investigated which aperture yields better signal-to-noise measurements.  We found again that at the faintest magnitudes, the small elliptical apertures tend to do better, with signal-to-noise ratios $\sim$ 10\% higher at H$_{160} =$ 29 mag, thus we adopt the fluxes measured in these apertures for our analysis.  To correct the flux in each band to the total flux, we obtained the H$_{160}$-band flux in a larger Kron aperture with PHOT\_AUTOPARAMS values of 2.5, 3.5 (the standard SExtractor MAG\_AUTO), and then use the ratio of this flux to the H$_{160}$-band flux in the smaller elliptical aperture as an aperture correction which is applied to the fluxes in each band.

\subsection{Photometric Redshifts}
We use the EAZY software package \citep{brammer08} to measure the photometric redshifts for objects in each of our four fields.  EAZY uses non-negative linear combinations of templates to solve for the best-fitting photometric redshift.  EAZY also returns the $\chi^2$ distribution, from which we can construct the redshift probability distribution function, $P(z)$, defined as $P(z) \propto$ exp($-\chi^2/2$), normalized such that the integral of $P(z)$ over all redshifts is unity.  We used the set of templates provided with EAZY based on the P\'{E}GASE stellar population synthesis models \citep{fioc97}.  We also include an additional template based on the rest-frame ultraviolet (UV) spectrum of the young, unreddened galaxy BX418 \citep{erb10}, as it retains characteristics expected in high-redshift galaxies, such as strong \lya\ emission.  EAZY assumes the IGM prescription of \citet{madau95}.  We note that as the luminosity functions at very high redshift are not yet very well-constrained, we elected to not include magnitude priors.

\subsection{Sample Selection}

Our goal is to study the evolution of the rest-UV color from $z =$ 4 -- 8.  To do this, we now construct five independent galaxy samples at each redshift of $z_{sample} =$ 4 (3.5 $\lesssim z \lesssim$ 4.5), $z_{sample} =$ 5 (4.5 $\lesssim z \lesssim$ 5.5), $z_{sample} =$ 6 (5.5 $\lesssim z \lesssim$ 6.5), $z_{sample} =$ 7 (6.5 $\lesssim z \lesssim$ 7.5) and $z_{sample} =$ 8 (7.5 $\lesssim z \lesssim$ 8.5).  We do not base these samples solely on the best-fit photometric redshift; rather we use the information available in the full $P(z)$ function.  We do this by using the following selection criteria:
\vspace{2mm}

\indent 1) Signal-to-noise in both J$_{125}$ and H$_{160}$ $\geq$ 3.5\\
\vspace{-2mm}

\indent 2) $\mathcal{P}_{primary}$ $>$ 0.7\\
\vspace{-2mm}

\indent 3) $\mathcal{P}_{z_{sample}}$ $>$ 0.25\\
\vspace{-2mm}

\indent 4) $z_{sample}$ - 1 $\leq$ z$_{phot}$ $\leq$ $z_{sample}$ + 1\\
\vspace{-2mm}

\indent 5) $\chi^2_{EAZY} \leq$ 10\\

\noindent where $\mathcal{P}_{primary}$ is the integral of $P(z)$ in the primary redshift peak, and $\mathcal{P}_{z_{sample}}$ is the integral of $P(z)$ in the redshift range of the sample of interest (i.e., $z_{sample} -$ 0.5 to $z_{sample} +$ 0.5).

\begin{figure*}[!ht]
\epsscale{1.15}
\plottwo{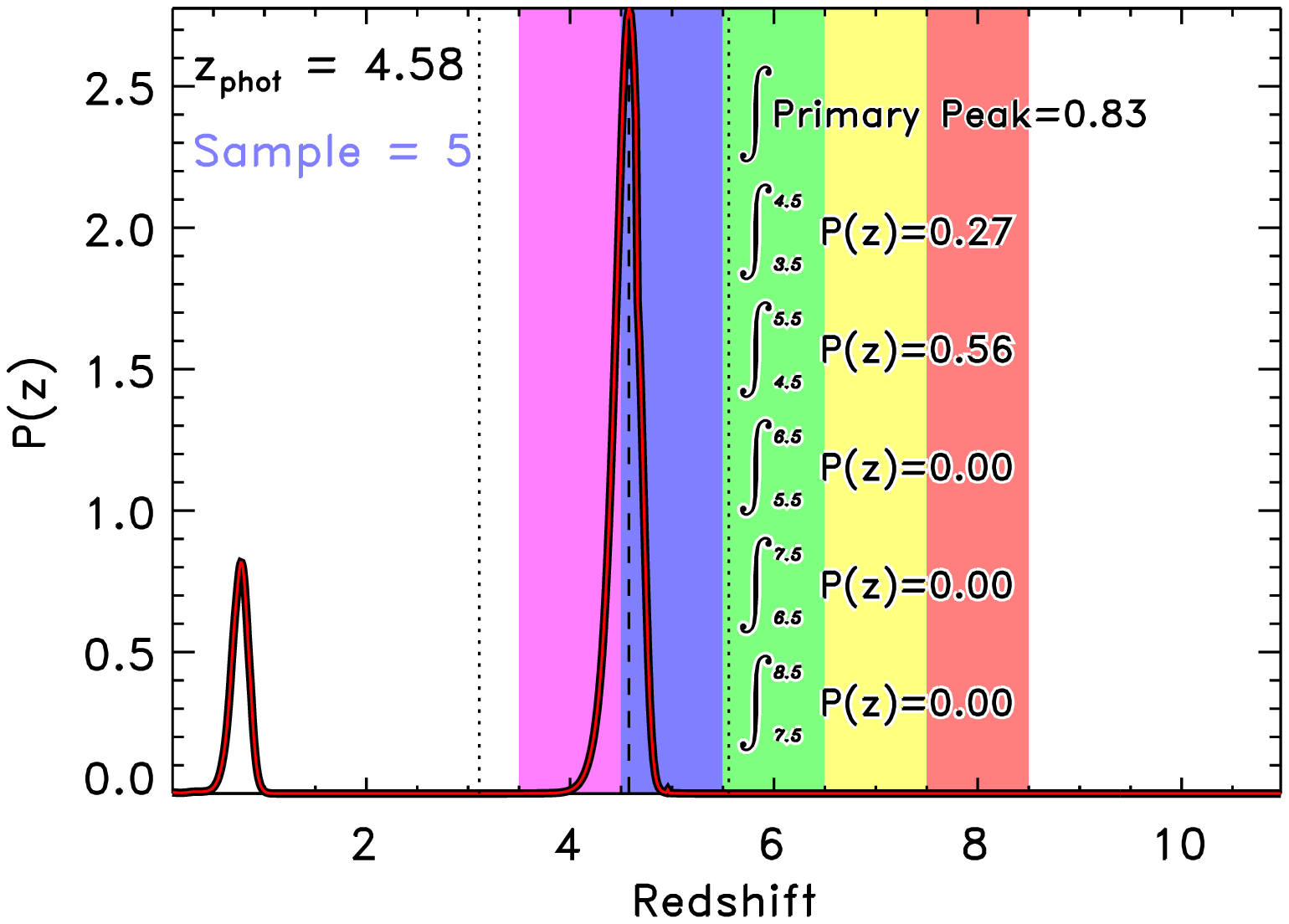}{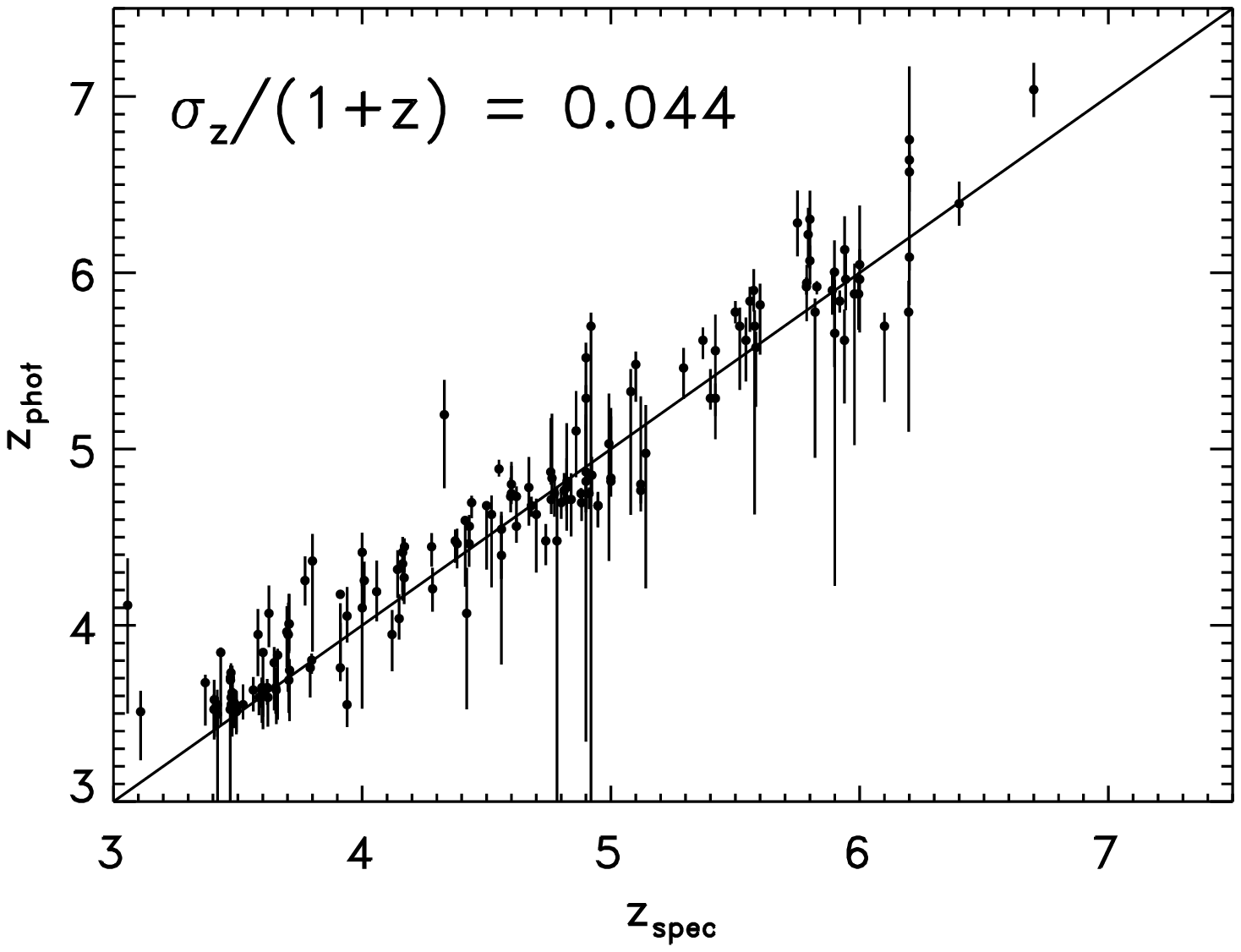}
\vspace{-2mm}
\caption{Left) An example of our sample selection.  This plot shows the $P(z)$ curve for an object in our $z =$ 5 sample.  The best-fit redshift for this object is $z_{phot} =$ 4.58, and it contains $>$ 80\% of the area under the $P(z)$ curve in the primary redshift peak.  Over 50\% of the area under the $P(z)$ curve falls in the 4.5 $< z <$ 5.5 (i.e., $z =$ 5) region, while only $\sim$25\% of the area falls in the $z =$ 4 region, thus it is placed in the $z =$ 5 sample.  Right) A comparison between the spectroscopic and photometric redshifts for the 157 galaxies in our sample with available spectroscopy.  After excluding 12 outliers (not shown), we find excellent agreement, with $\sigma_{z}/(1+z) =$ 0.044.}
\end{figure*} 

The first criterion ensures that the object is significantly detected in two bands, and is thus unlikely to be spurious.  The second criterion demands that the primary solution be a dominant one, in that only $<$ 30\% of the area under the $P(z)$ curve can be in a secondary solution (see \S 3.4 for the effects of varying this criterion).  The third criterion ensures that sufficient area under the $P(z)$ curve resides in the regime of a given sample (i.e., $z$ = 3.5 -- 4.5 for the $z =$ 4 sample), while the fourth criterion ensures that the best-fit photometric redshift is near to the sample redshift as well.  In practice, the third and fourth criteria individually don't exclude many sources, but the combination of the two, combined with the second criterion, ensures a clean sample.  The final criterion ensures that EAZY is providing a reasonable fit; this final criterion does not exclude many sources.  Finally, to ensure that objects only show up in one sample, if an object satisfies these criteria for multiple samples, then it is placed in the sample where the majority of the area under its $P(z)$ curve resides (i.e., if $z_{phot} =$ 6.5, and $\mathcal{P}_7$ >  $\mathcal{P}_6$, then this object would be placed in the $z =$ 7 sample).  We highlight an example of this process in Figure 1.  We emphasize that at these redshifts, while the $P(z)$ does primarily key off of the Lyman break, it includes information from all filters, allowing the construction of an accurate $P(z)$ function.

After running EAZY on the catalogs for each of our four fields, we select galaxies in each of our five redshift bins in each field using the above criteria.  We visually inspected each source, to ensure that it is real, excluding interlopers such as diffraction spikes or oversplit parts of bright nearby galaxies.  We also checked for duplicates when a galaxy was in an overlap region between any of our four fields; in these cases, we kept the object in the sample that had the deeper photometry.  Our final sample comprises 2812 objects, 113 of which are in our $z =$ 7 and 8 samples (i.e., $z \gtrsim$ 6.5).  To test the validity of our sample selection, we compared the photometric redshifts of objects in our sample to known spectroscopic redshifts in GOODS-S \citep[e.g.][and references therein]{grazian06, wuyts08, vanzella08}, finding excellent agreement, and very few catastrophic outliers.  In summary, we found spectroscopic redshift matches for 157 of our galaxies.  Of these, we found that 12 were outliers, which we defined as having a difference between the spectroscopic and photometric redshift of $\Delta z$ $>$ 0.5, and that the spectroscopic redshift was not consistent within 3$\sigma$ of the photometric redshift.  All of these 12 had $z_{phot}$ $\sim$ 4, with 11 being lower redshift galaxies ($z_{spec} <$ 2), and one being at higher redshift ($z_{spec} =$ 5.1).  We removed these 12 outliers from our sample.  We found that once these outliers were removed, our photometric redshift accuracy was quite good, with $\sigma_{z}/(1+z) =$ 0.044.  We note that only 10 galaxies in our sample with $z >$ 6 have spectra, thus there is clearly a need for larger spectroscopic samples at the highest redshifts.  The comparison between photometric and spectroscopic redshifts for our cleaned sample can be found in Figure 1, and the details of our full sample are summarized in Table 2.  For consistency, we assume the photometric redshift in the analysis below even for those objects with confirmed spectroscopic redshifts.

\begin{deluxetable}{cccccc}
\tabletypesize{\small}
\tablecaption{Galaxy Sample}
\tablewidth{0.45\textwidth}
\tablehead{
\colhead{Field} & \colhead{$N_{z=4}$} & \colhead{$N_{z=5}$} & \colhead{$N_{z=6}$} & \colhead{$N_{z=7}$}  & \colhead{$N_{z=8}$}\\
}
\startdata
HUDF&177&71&50&26&14\\
GOODS-S DEEP&846&231&89&19&13\\
GOODS-S WIDE&265&68&23&3&1\\
GOODS-S ERS&647&171&61&32&5\\
\hline
Total Sample&1935&541&223&80&33
\enddata
\tablecomments{A summary of the number of candidate galaxies in each redshift
  bin from each dataset.  There are a total of 2812 candidate galaxies in our sample.}
\end{deluxetable}

We note that a few of the F105W visits in the HUDF are affected by persistence from a previously imaged globular cluster.  The affected regions are a relatively small portion of the image, thus we have included all visits in our final stacked F105W image to obtain the maximum depth possible.  To see if any objects in our final sample are affected, we performed photometry on an image incorporating the subset of $\sim$ 75\% of the F105W exposures unaffected by strong persistence.  We find the difference between the fluxes of objects in our sample between the two images is small, with a scatter of $\sim$ 12\% for galaxies with Y$_{105}$ < 29 mag.  We note that much of this scatter is likely due to the decreased depth in the persistence-free image.  No objects had F105W fluxes which differed by $>$ 3$\sigma$, and only three objects had F105W fluxes differing by $>$ 2$\sigma$ (two of which were near to the image edge), thus we conclude that the persistence is not having a significant effect on the F105W fluxes of the HUDF objects in our sample.

\begin{figure*}[!ht]
\epsscale{1.1}
\plottwo{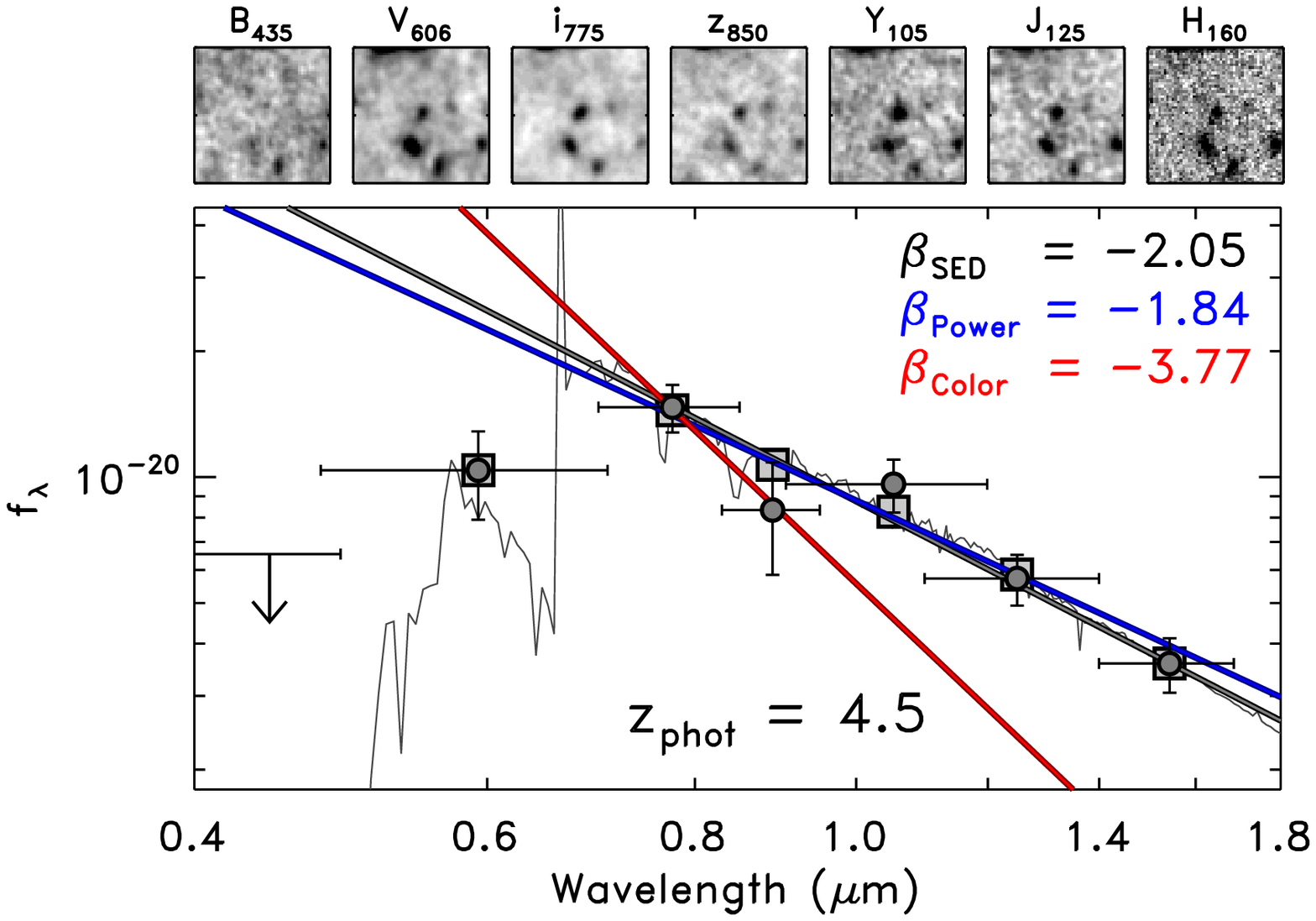}{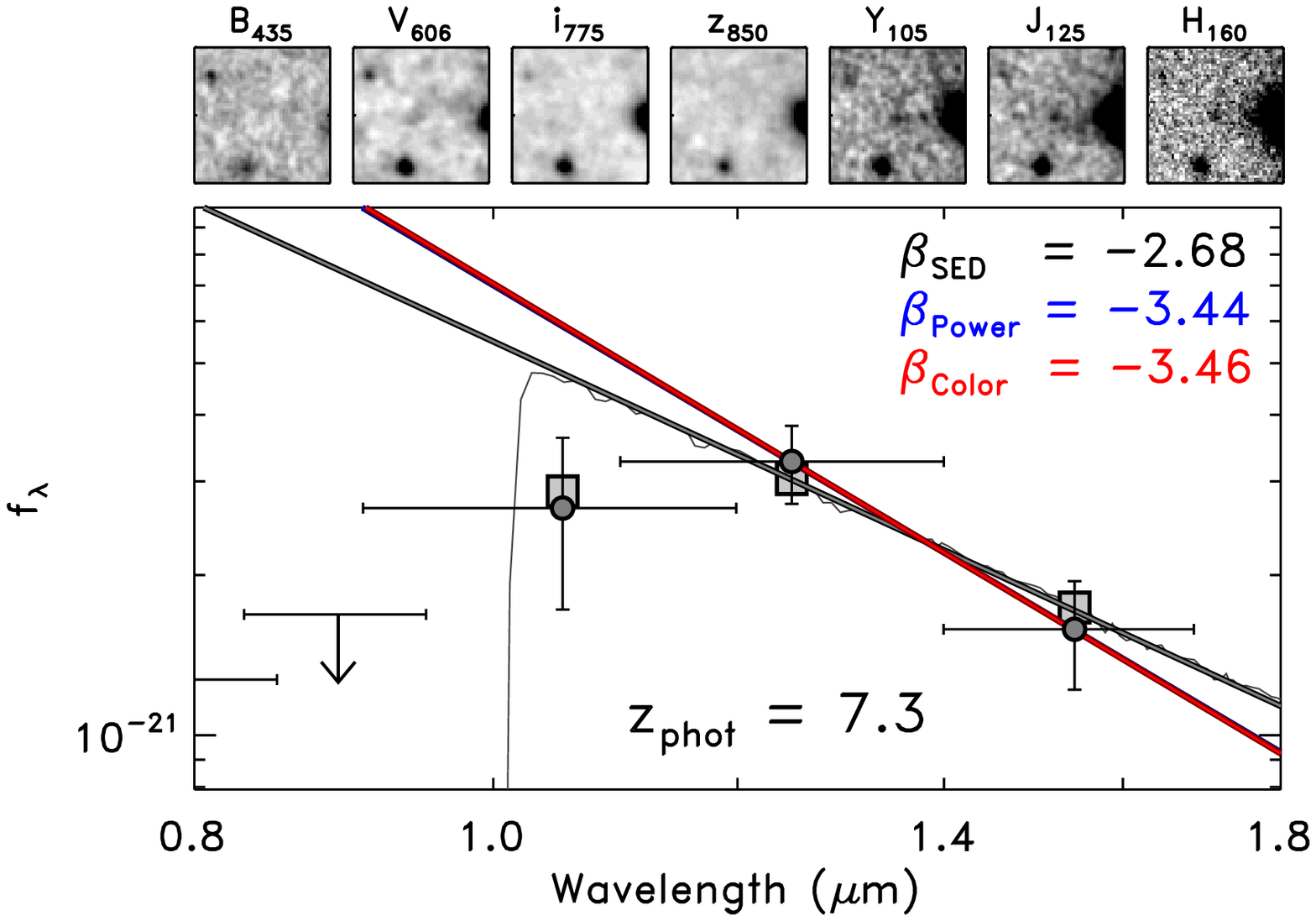}
\plottwo{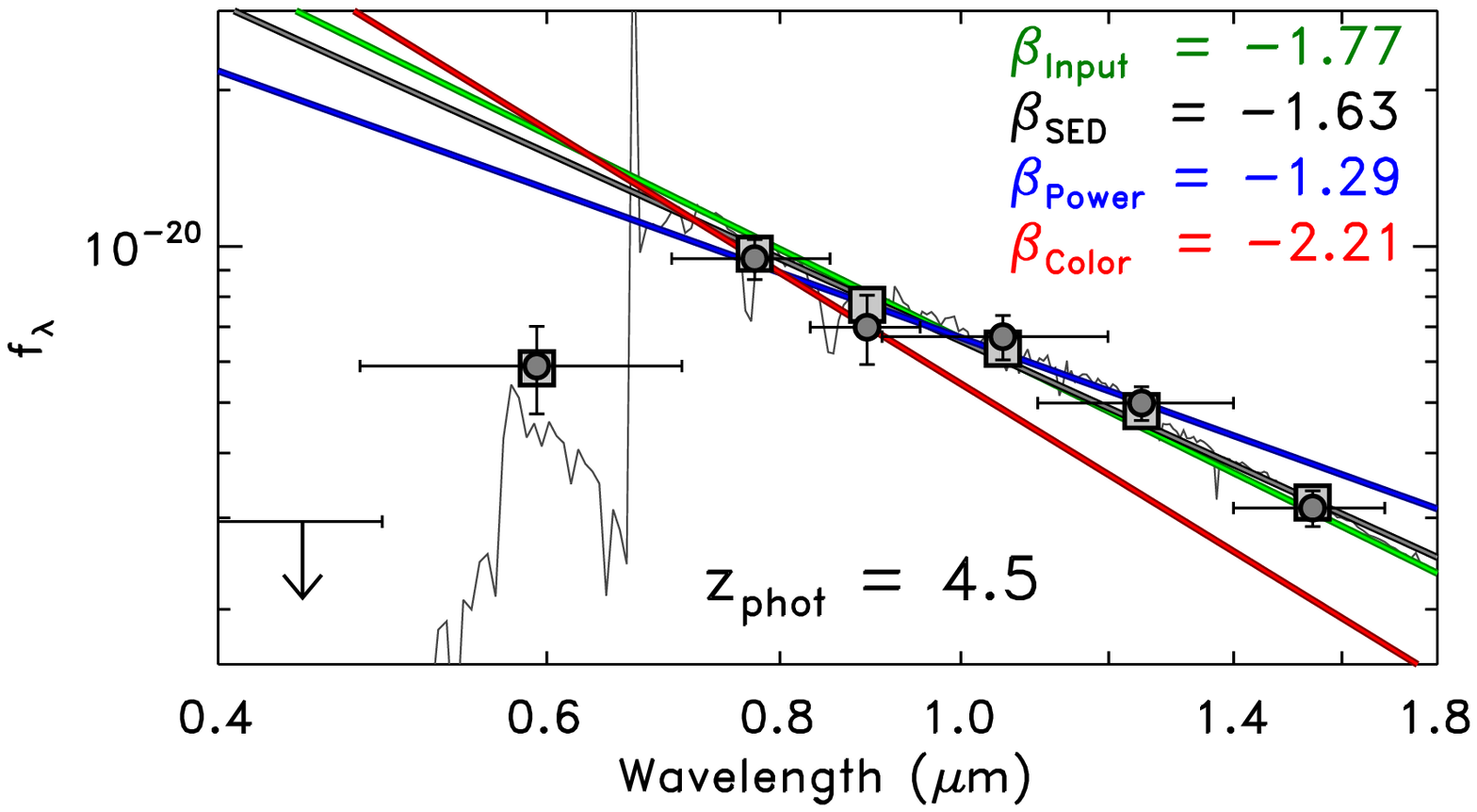}{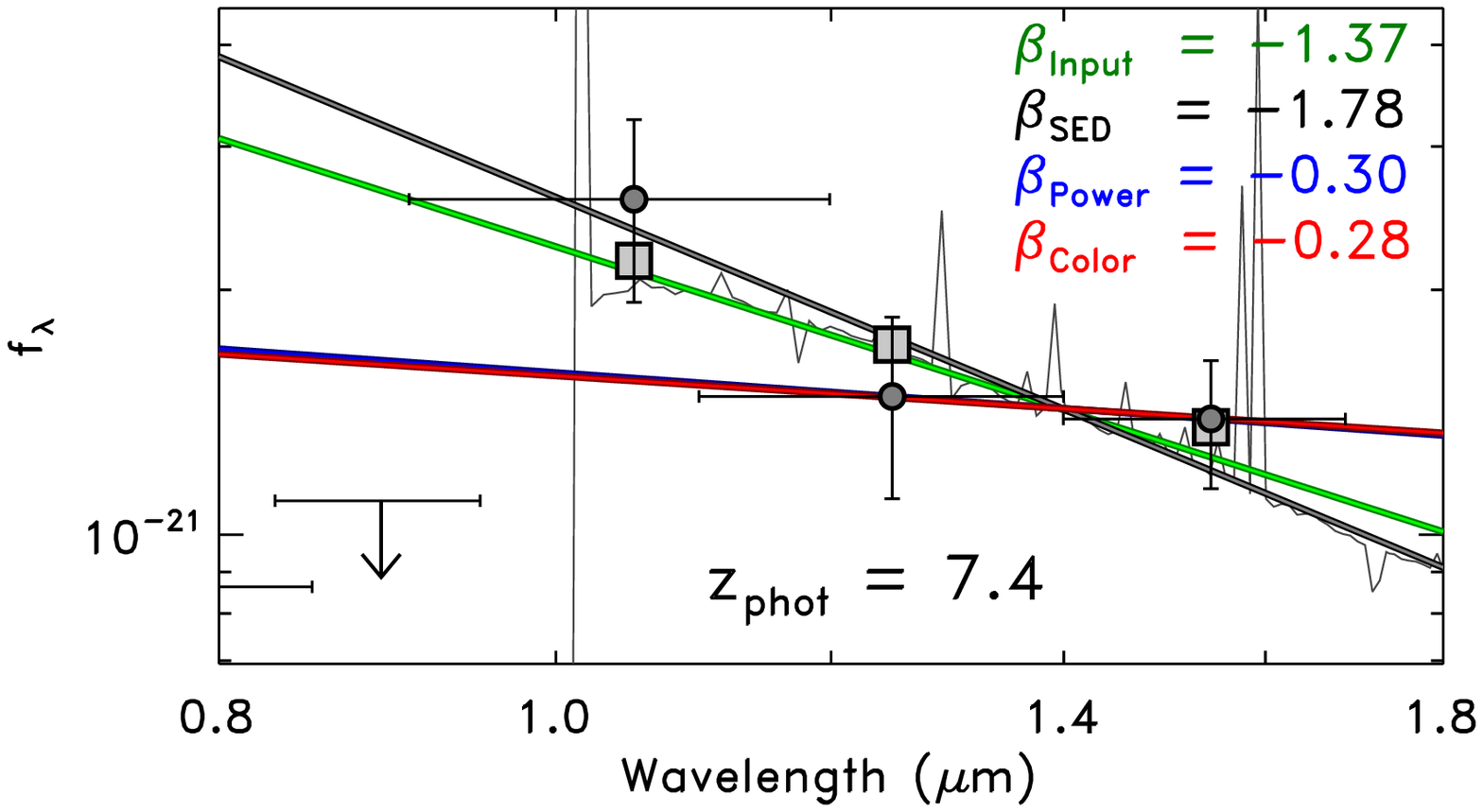}
\caption{Top row: Example objects from our $z =$ 4 sample (left) and $z =$ 7 sample (right).  The top images show 3\arcs\ stamps of each object in the {\it HST} bands.  The plots show the best-fit model spectrum, as well as the UV spectral slopes from the SED-fitting method (gray), power-law method (blue) and single-color method (red).  The circles show the measured fluxes of the objects in each band, while the squares show the bandpass-averaged fluxes from the best-fit model.  Bottom row: Simulated objects at $z =$ 4 (left) and $z =$ 7 (right), each chosen to have a photometric redshift and H$_{160}$ magnitude similar to the real object in the top panel.  In all four cases, variations in the photometry result in a single-color-derived value of $\beta$ which is different from the other two methods (and the true value).  In general at $z =$ 4, the SED-fitting and power-law methods both work fairly well, but by $z =$ 7, the SED-fitting method appears more reliable, as the power-law method is reliant upon a single color.}
\end{figure*} 

\vspace{10mm}

\section{Ultraviolet Spectral Slope}

The rest-frame ultraviolet colors of galaxies can be parameterized by the UV spectral slope, $\beta$, defined in \citet{calzetti94} as f$_{\lambda}$ $\propto$ $\lambda^{\beta}$.  This parameter was designed to be measured spectroscopically, in specific windows defined by \citet{calzetti94} to omit spectral emission and absorption features.  However, continuum spectroscopy is extremely difficult at $z \sim$ 4, and currently impossible at $z \sim$ 7.  Thus, $\beta$ is commonly estimated at high redshift from a single rest-frame UV color \citep[e.g.,][]{meurer97, hathi08, bouwens10b}, where $\beta$ is proportional to a single color over a redshift range of $\Delta z$ $\sim$ 1.  However, with our dataset we have multiple rest-frame UV colors over our whole sample except for the $z =$ 8 bin.  For example, at $z \sim$ 4, there are six photometric bands redward of the Lyman break, covering rest-frame wavelengths from $\sim$ 1000 -- 3500 \AA.  Making use of all of the available photometric bands should yield more robust estimates of $\beta$, as it will be less susceptible to one outlying data point.  

\subsection{Measuring $\beta$}
We measure $\beta$ by performing spectral energy distribution (SED) fitting on our sample of galaxies, finding the best-fitting synthetic stellar population using updated (2011) models of \citet{bruzual03}. We created a suite of stellar population models assuming a Salpeter initial mass function (IMF), varying the metallicity (0.02 $\leq Z \leq$ 1 $Z$\sol), age (1 Myr $\leq$ t $\leq$ t$_{H}$), dust extinction (0 $\leq$ A$_{V}$ $\leq$ 2 mag, using the attenuation law of \citet{calzetti00}) and star-formation history (SFH $\propto$ exp$^{-t/\tau}$, with $\tau$ = 10$^{5}$, 10$^{7}$, 10$^{8}$, 10$^{9}$, 10$^{10}$, 10$^{11}$, $-$3 $\times$ 10$^{8}$, $-$10$^{9}$ and $-$10$^{10}$ yr).  Note the negative sign of the last three values of $\tau$; these indicate rising star-formation histories, which mounting observational \citep[e.g.,][]{stark09,papovich11, lee11} and theoretical \citep[e.g.,][]{finlator11} evidence indicates is the dominant mode at $z \gtrsim$ 3.  This procedure is similar to that used in \citet{finkelstein10a}, thus the reader is referred there for more details.

We include nebular emission lines, adding hydrogen lines in the Lyman, Balmer, Paschen and Bracket series, where the line strengths are directly proportional to the number of ionizing photons (which is dependent on the population age).  Metal lines are also added, where the metal line ratios are a function of both the number of ionizing photons and the metallicity.  We used the Lyman, Balmer and metal line ratios from \citet{inoue11}, while the Paschen and Bracket line ratios come from \citet{osterbrock06}.  All emission line strengths are dependent on the ionizing escape fraction, which for simplicity we assume is zero (and is likely to be $<$ 50\% in the galaxies we observed \citep[e.g.,][]{yajima11}).  Further details on the nebular line calculations will be available in Salmon et al.\ (in prep).

\begin{deluxetable*}{cccccccc}
\tabletypesize{\small}
\tablecaption{Properties of our Galaxy Sample}
\tablewidth{0pt}
\tablehead{
\colhead{ID} & \colhead{Sample} & \colhead{RA} & \colhead{Dec} & \colhead{z$_{phot}$} & \colhead{$\beta$} & \colhead{$M_{1500}$} & \colhead{log Stellar Mass}\\
\colhead{$ $} & \colhead{$ $} & \colhead{(J2000)} & \colhead{(J2000)} & \colhead{(68\% Conf. Range)} & \colhead{(68\% Conf. Range)} & \colhead{(68\% Conf. Range)} & \colhead{(68\% Conf. Range)}\\
}
\startdata
HUDF\_339&z4& 53.154957&$-$27.764288&3.42 (3.31 -- 3.73)&$-$2.29 ($-$2.58 -- $-$2.17)&$-$17.50 ($-$17.70 -- $-$17.39)& 7.73 ( 7.30 -- 7.95)\\
DEEP\_16412&z5& 53.047127&$-$27.807669&4.48 (4.22 -- 4.71)&$-$2.38 ($-$2.41 -- $-$1.86)&$-$19.37 ($-$19.41 -- $-$19.20)& 8.45 ( 8.42 -- 9.38)\\
WIDE\_13924&z5& 53.168690&$-$27.857683&4.53 (4.39 -- 4.66)&$-$2.00 ($-$2.28 -- $-$1.83)&$-$20.55 ($-$20.58 -- $-$20.41)& 9.14 ( 8.95 -- 9.61)\\
ERS\_4619&z5& 53.158905&$-$27.688990&4.46 (4.09 -- 4.85)&$-$2.41 ($-$2.68 -- $-$2.07)&$-$18.73 ($-$18.93 -- $-$18.52)& 8.45 ( 7.49 -- 8.44)\\
HUDF\_3233&z6& 53.178478&$-$27.788225&5.64 (5.54 -- 5.71)&$-$1.95 ($-$2.01 -- $-$1.77)&$-$19.77 ($-$19.81 -- $-$19.70)& 8.88 ( 8.57 -- 9.76)\\
WIDE\_14210&z6& 53.225220&$-$27.853327&5.54 (5.44 -- 5.78)&$-$2.39 ($-$2.61 -- $-$1.91)&$-$20.27 ($-$20.39 -- $-$20.18)& 8.56 ( 8.37 -- 9.62)\\
ERS\_20668&z6& 53.069393&$-$27.746489&5.56 (5.44 -- 5.64)&$-$2.51 ($-$2.54 -- $-$2.29)&$-$20.76 ($-$20.77 -- $-$20.67)& 9.24 ( 8.47 -- 9.41)\\
WIDE\_8994&z7& 53.239098&$-$27.889389&6.62 (6.34 -- 6.93)&$-$2.37 ($-$2.68 -- $-$1.50)&$-$20.00 ($-$20.18 -- $-$19.85)& 9.06 ( 8.25 -- 9.55)\\
HUDF\_1838&z8& 53.186256&$-$27.778973&7.56 (7.34 -- 7.79)&$-$2.43 ($-$2.56 -- $-$2.02)&$-$19.53 ($-$19.66 -- $-$19.51)& 8.18 ( 8.09 -- 9.25)\\
ERS\_14661&z8& 53.147747&$-$27.723196&7.64 (4.75 -- 8.62)&$-$2.45 ($-$2.92 -- $-$1.62)&$-$19.79 ($-$20.03 -- $-$19.47)& 8.39 ( 7.86 -- 9.78)\\
\enddata
\tablecomments{A catalog of our 2812 $z =$ 4 -- 8 galaxy candidates, with their
derived properties.  The values in parenthesis represent the 68\%
confidence range on the derived parameters.  Here we show 10
representative galaxies from our four fields and five redshift bins; the full table is available in
machine-readable format in the electronic version.}
\end{deluxetable*}

We find the best-fit model via $\chi^2$ minimization, and we measure the value of $\beta$ directly from this best-fit spectrum by fitting a power law to the spectrum using the wavelength windows specified by \citet{calzetti94}.  We also measure the absolute magnitude at 1500 \AA\ (M$_{1500}$) from this spectrum by integrating the flux within a 100 \AA-wide square bandpass centered on 1500 \AA.  We show an example of the results from this procedure in the top panels of Figure 2.  We then estimate the uncertainty on $\beta$ by running a series of Monte Carlo simulations, varying each object's photometry by an amount proportional to the photometric uncertainty, and re-deriving a best-fit model.  In each simulation, we re-measure $\beta$ in the same way as above, and then we use the distribution of $\beta$'s to calculate the 68\% confidence range.  This method has several advantages.  First, provided the choice of stellar population models is appropriate and one can obtain a good fit, this method uses all the available colors to compute an accurate value of $\beta$, rather than a single color which is much more subject to photometric outliers.  Secondly, this method obtains an accurate value of $\beta$ for the specific photometric redshift.  The single color method is designed to work over $\Delta z$ $\sim$ 1, while the 68\% confidence ranges on our photometric redshifts are typically $\Delta z$ $\sim$ 0.3, thus fitting $\beta$ at the best-fit photometric redshift should yield a more accurate result.  In addition, the uncertainty in the redshift is accounted for in the uncertainty on $\beta$, as in our Monte Carlo simulations we draw a new value of the redshift from the galaxy's $P(z)$ curve in each simulation (setting $P(z < z_{phot}-2) = 0$).  There is a disadvantage to this method as well, in that one is restricted to the range of $\beta$ allowable in the models one chooses.  In particular, for our suite of models, the bluest value of $\beta$ is $-$3.15; if a galaxy had a UV spectral slope bluer than this, it would not be accurately recovered.  However, as we show below, we do not find any galaxies with $\beta$ $<$ $-$3 (the bluest galaxy in our sample has a measured value of $\beta = -$2.91), thus we have not yet reached this limiting parameter value.  Future studies probing closer to the epoch of first star formation will need to include models more appropriate to those expected stellar populations.

We reiterate that the choice of models should have no impact on the derived values of $\beta$.  Provided the model is a good fit to the data, two models with differing ages or star-formation histories may have similar UV spectral slopes, and thus have similar measured values of $\beta$.  To be sure this is the case, we eliminate galaxies with poor SED fits, defined as reduced $\chi^2$ $>$ 10, from our analysis below.  We find this rejects very few galaxies; only 9 out of our total sample of 2812 were excluded due to this cut.  Comparing the distribution of $\beta$ values from galaxies excluded by this cut to the full sample, we see a peak at similar values ($\beta \sim -$2), while the excluded galaxies do appear to have an over-abundance of very red ($\beta >$ 0) colors.  Had we chosen $\chi^2$ $>$ 5 for the cut, we would not have excluded many more anomalously red galaxies, thus we move forward with our original choice of $\chi^2$ $>$ 10.  In either case, the conclusions below are unaffected by this choice.  

We note that since we do use all photometric bands, including those affected by the Lyman break and \lya\ emission, it is possible that our treatment of these two properties results in a slight systematic offset of the results.  However, given the general agreement between our method, and the power-law method (where one fits a power-law to all photometry redward of the break; see \S 3.2), such an offset is likely minimal.  We check this by comparing our best-fit $\beta$ values to those obtained via fitting a power-law to the photometry.  We find these values are consistent within their scatter, with a typical difference between the two methods of $\Delta\beta =$ 0.09 $\pm$ 0.20 at $z =$ 5 and $\Delta\beta =$ 0.25 $\pm$ 0.40 at $z =$ 7.  We have also checked via our simulations (see \S 3.2) whether significant differences arise when excluding the Lyman-break bands from the fit at $z =$ 4 and 5; we find essentially no difference for galaxies with $\beta > -$2, while bluer galaxies tend to be measured with $\beta$ only 0.05 bluer when excluding the band around the break.  We conclude that our inclusion of the Lyman break-affected bands are not significantly affecting our results.  Finally, we also tested how well we recover the simulated $z =$ 4 galaxy values of $\beta$ when using only the ACS photometry, which is analogous to the case at $z =$ 7, where galaxies are only detected in three bands.  We find that, as expected, the scatter in $\beta$ increases by $\sim$ 0.1 and 0.2 at H$_{160} =$ 24 and 29, respectively, similar to that found in the $z =$ 7 simulations below.  The difference in the recovered versus input values of $\beta$ (i.e., the bias) are less than the scatter, at $\Delta\beta <$ 0.05 for bright galaxies, and $\Delta\beta <$ 0.15 for faint galaxies.  We ran a similar test using only the B$_{435}$V$_{606}$i$_{775}$ photometry, which is analogous to the case at $z =$ 8, and found an increase in the scatter on $\beta$ of 0.5 at all magnitudes.  The bias was larger as well, with $\Delta\beta >$ 0.5 for galaxies fainter than 28th magnitude.  This shows that the current results at $z =$ 8 we discuss are preliminary, due to the single detected color available, which may be contaminated by both the Lyman break and \lya\ emission.  In Table 3 we list the celestial coordinates and derived values of redshift, stellar mass, M$_{1500}$ and $\beta$ for all objects in our sample.

\begin{figure*}[!t]
\epsscale{0.9}
\plotone{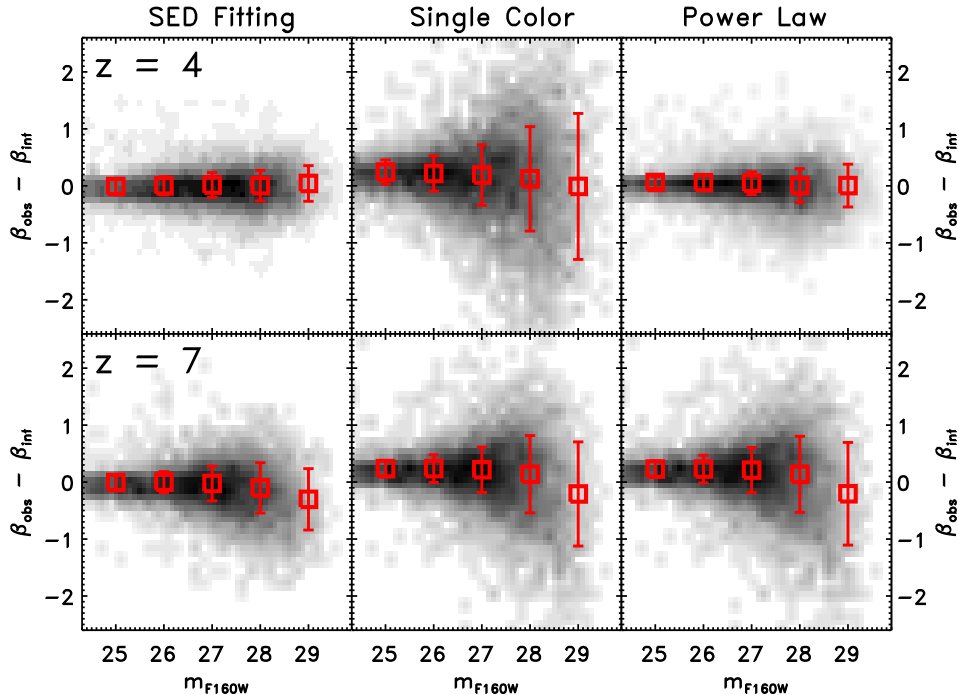}
\vspace{-2mm}
\caption{Results from our simulations at $z =$ 4 and $z =$ 7 (top and bottom panels, respectively) in the HUDF.  In these simulations, we create artificial galaxies and insert them into the data.  The galaxies are then photometered, and samples at different redshifts are then chosen in the same way as done on the actual data.  All panels show the difference between the input value of the UV spectral slope $\beta$ and that recovered from the photometry as a function of input H$_{160}$ magnitude.  The left panel of both rows shows the results when $\beta$ is recovered during the SED fitting process, which uses all available photometric bands.  The middle panel shows the results when using only a single rest-UV color.  The right panel shows the results when fitting a power law as in \citet{bouwens12}.  The shading denotes the density of simulated galaxies, while the red squares denote the mean difference between simulated and observed values of $\beta$ in bins of $\Delta$m $=$ 1.  We have run similar simulations in the CANDELS fields which also show that the SED fitting method minimizes the scatter in $\beta$.}
\end{figure*} 

\vspace{5mm}

\subsection{Testing the Method}
To test our SED-fitting method of measuring $\beta$, we have run a series of simulations using the data in the HUDF.  In these simulations, we created mock galaxies with the GALFIT software \citep{peng02}, using a range of half-light radii, axial ratios and Sersic indices.  We assumed a distribution of H$_{160}$ magnitudes that is Gaussian at bright magnitudes and log-normal at faint magnitudes (which reproduces the observed number counts, extending the trend to fainter magnitudes), and we created stellar population models with a range of ages, metallicities and dust extinctions to derive the fluxes in the remaining photometric bands.  The input distribution of half-light radii and stellar population properties were chosen such that the output radius and $J-H$ color distributions were closely matched to those of the galaxy samples.  These ``mock'' galaxies were then inserted into the appropriate image.  As the HUDF is only a single WFC3 field, to avoid overcrowding in the simulations we performed twenty iterations, each time inserting 1000 galaxies (forcing the positions of the simulated galaxies to avoid those of real objects in the image using the SExtractor segmentation map).  We performed these simulations separately for each redshift bin, assuming a flat distribution of input redshifts.

These images were then photometered using SExtractor in the same manner as was done on the original data, and photometric redshifts were then computed using EAZY.  Using the recovered photometry and photometric redshifts of these mock galaxies, we selected out redshift samples in each of the simulations using the same five criteria outlined in \S 2.4.  In this way, the recovered galaxies from the simulations best match the properties of the galaxies in our sample.  We then measured the value of $\beta$ in three ways; first, we performed SED fitting on these mock galaxies using the recovered photometry in the same way as on our object samples.  Secondly, we measured $\beta$ using a single color, using the relations from \citet{hathi08} for the $z =$ 4 sample, that those from \citet{dunlop12} for the $z =$ 5 -- 8 samples (using the Y$_{105} - $J$_{125}$ color for $z =$ 5 and 6, and the  J$_{125} - $H$_{160}$ color for $z =$ 7 and 8).  Thirdly, we measured $\beta$ via a power-law fit to the available bandpasses, using the same bandpasses in each redshift bin as done in \citet[][see also Castellano et al.\ 2012]{bouwens12}.  Examples of simulated galaxies are shown in the bottom panels of Figure 2.

\begin{figure*}[!t]
\epsscale{1.1}
\plottwo{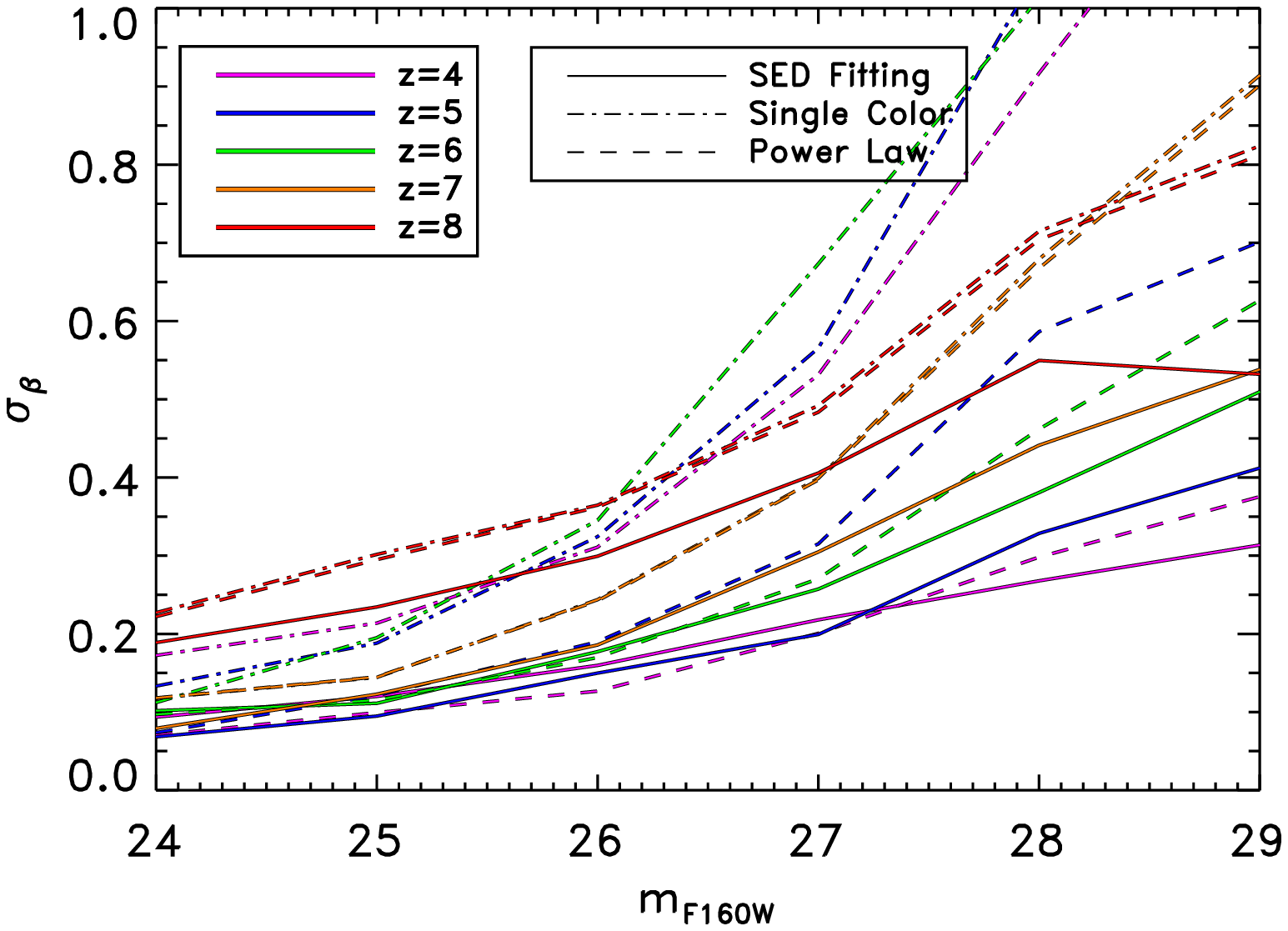}{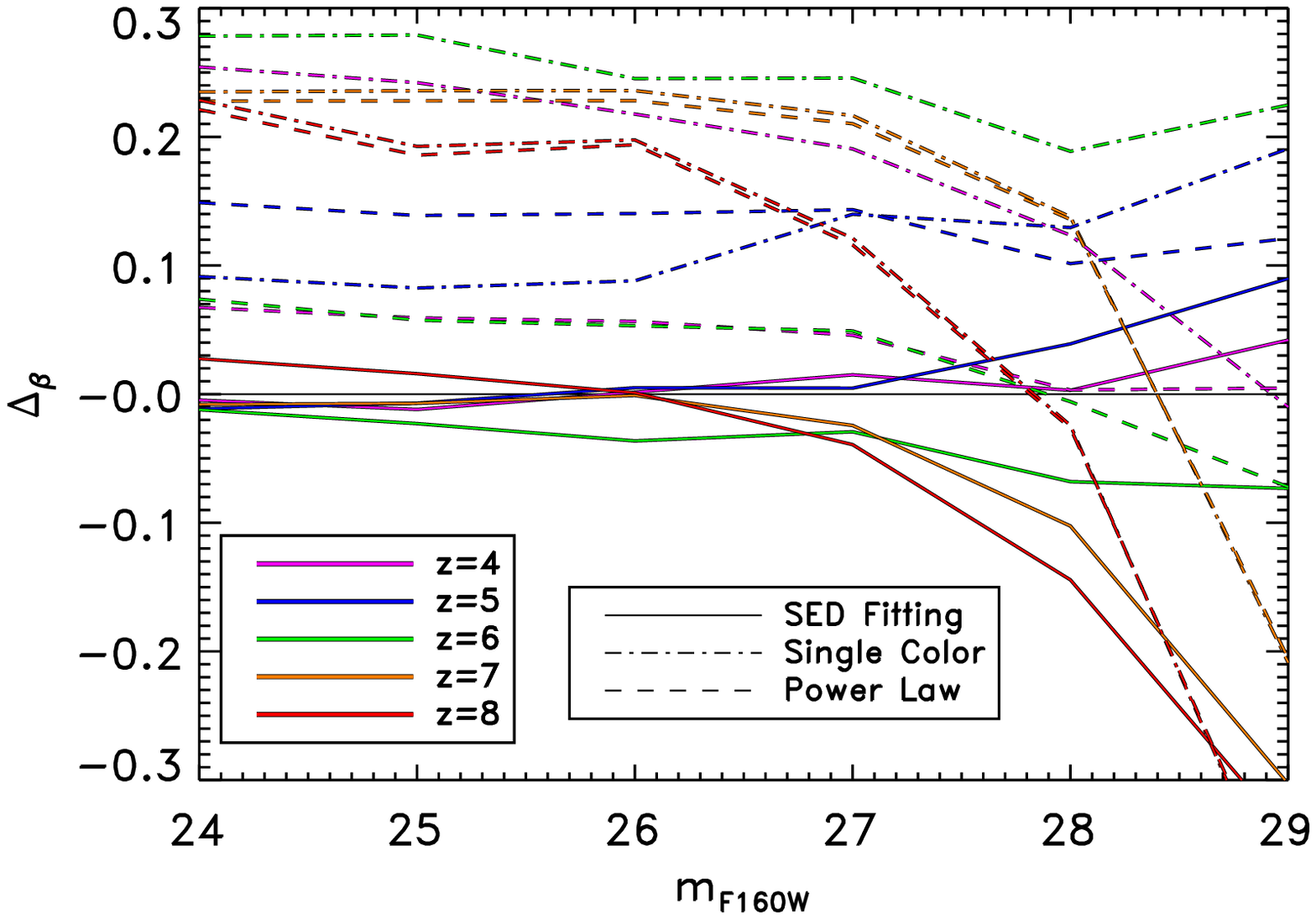}
\vspace{-2mm}
\caption{Left) The recovered uncertainty in $\beta$ from the SED fitting method (solid lines), the single-color method (dashed lines), and the power-law method (dotted lines) from our simulations in the HUDF (we expect the observed trends to shift to brighter magnitudes in the shallower datasets).  At all redshifts and at all magnitudes, measuring $\beta$ during the SED fitting process yields a smaller scatter than using only a single color.  The SED-fitting method yields comparable results to the power-law method at $z \leq$ 6, but it results in a much smaller scatter at $z =$ 7 and 8.  Right)  The photometric bias from our simulations.  The SED fitting method has no bias at all but the faintest magnitudes, while the single color and power-law methods suffer a red bias at bright magnitudes, transitioning to a blue bias at faint magnitudes.}
\end{figure*}

The results from these simulations in the HUDF are shown in Figure 3, where the shading denotes the density of simulated galaxies.  In the top row, we show the results from our $z =$ 4 simulations, plotting the difference between the input value of $\beta$ and the recovered value from the SED-fitting method in the left panel, the difference in $\beta$ from the single-color method in the middle panel, and that from the power-law method in the right-hand panel.  The bottom row shows the same three methods for our $z =$ 7 simulations.  In each panel, we compute the median difference (i.e., bias) as well as the scatter in bins of $\Delta$m$_{H}$ = 1 mag, shown as the red squares and error bars in both panels.  Two items are immediately obvious; 1) Both the SED-fitting and power-law methods result in a smaller scatter at all magnitudes than the single-color method, and 2) the number of galaxies ``catastrophically'' scattered to $\Delta\beta > \pm$1 is much less when using the SED-fitting or power-law methods.  We note that at $z =$ 7 the power-law and single-color methods suffer the same drawbacks, as the power-law method excludes the Y-band due to the possible contribution by \lya\ emission.  Our SED-fitting method still makes use of the Y-band photometry, allowing \lya\ emission in the models to best-match the observed Y $-$ J$_{125}$ color, then excluding the spectral region around \lya\ when fitting $\beta$.  This appears to work well, as the scatter in the SED-fitting method at $z =$ 7 is less than the other two methods, and it does not show any bias due to poor fits to the \lya\ strength.  Similarly, this method allows us to report the first results for $\beta$ at $z =$ 8, which the other methods cannot do (though with a large scatter).

In Figure 4 we show the results of our simulations at all redshifts.  The left panel shows the photometric scatter, and it is apparent that the single-color method suffers a much larger scatter at all redshifts and magnitudes than the other two methods.  Additionally, while the SED-fitting and single-color methods yield similar results at $z =$ 4 -- 6, the SED-fitting method results in much less scatter at $z =$ 7 -- 8.  The right-hand panel shows the photometric bias.  At m$_{F160W} <$ 28, the SED-fitting method results in essentially no bias at all redshifts, while both the power-law and single-color methods result in a bias toward redder values of $\beta$.  This red bias is likely due to the fixed-redshift nature of these methods.  \citet{meurer99} show (in their Figure 3) that a correction needs to be made to the value of $\beta$ depending on the redshift of the galaxy; they find that for most redshifts, this correction is in the blueward direction, meaning that the galaxy is measured as redder than its true value, similar to what we find in our simulations.  Finally, we note that at $z =$ 7 -- 8, all three methods contain a blue bias at faint magnitudes (at least relative to the $z <$ 7 baseline), similar to the selection bias discussed in \citet{dunlop12}.  This bias originates from the fact that the same 2--3 colors used to measure $\beta$ are key to the selection of these galaxies.  For example, if the J$_{125}$-band flux of a galaxy fluctuates upward, it will be measured to have a bluer $\beta$ in our sample.  However, if the H$_{160}$-band flux experiences such a fluctuation, any measured $\beta$ would be redder, thus this galaxy is more likely to be assigned a lower photometric redshift, removing it from the sample.  This bias does appear to be somewhat smaller in the SED fitting method at the faintest magnitudes ($\Delta\beta$ $\sim$ $-$0.25) versus the other two ($\Delta\beta$ $\sim$ $-$0.5).  Given these simulation results, we conclude that the SED fitting method produces the most reliable estimate of $\beta$, minimizing both the photometric scatter and bias.  We thus adopt the value of $\beta$ obtained from the SED fitting method in our analysis below.

One possible complication to our simulations is that we have used the same set of models (i.e., the updated models of Bruzual \& Charlot) to create the simulated galaxies as we used to measure their recovered value of $\beta$, thus one may worry that the SED fitting method is biased to have a lower scatter.  To test this, we used a different set of models for the input colors of the simulated galaxies; we used the older 2003 version of the Bruzual \& Charlot spectral synthesis models, with a Milky Way \citep{cardelli89} extinction curve, while still using the updated synthesis models with the \citet{calzetti00} extinction curve when performing the SED fitting on the recovered objects.  We compared the recovered uncertainty in $\beta$ as a function of magnitude from this test to our baseline simulations, and found the scatter increased by $<$ 5\% (from H$_{160} =$ 24 -- 29) when using the different input assumptions.  We thus conclude that our choice of models is not biasing our results.

Finally, we note that these simulations also provide a secondary test to our photometric redshifts which is not model-dependent, as the templates we use with EAZY differ from those used to create the simulated galaxies.  We compared the recovered photometric redshift to the input value of the redshift in each of our simulations, and we find excellent agreement, with $\sigma_{z}/(1+z) =$ 0.02 -- 0.05.  The largest uncertainty of 0.05 occurs in the lowest redshift bin of $z =$ 4, likely due to the lack of a band completely blueward of the Lyman break for the lowest redshift galaxies in this bin.

\begin{figure*}[!th]
\epsscale{0.9}
\plotone{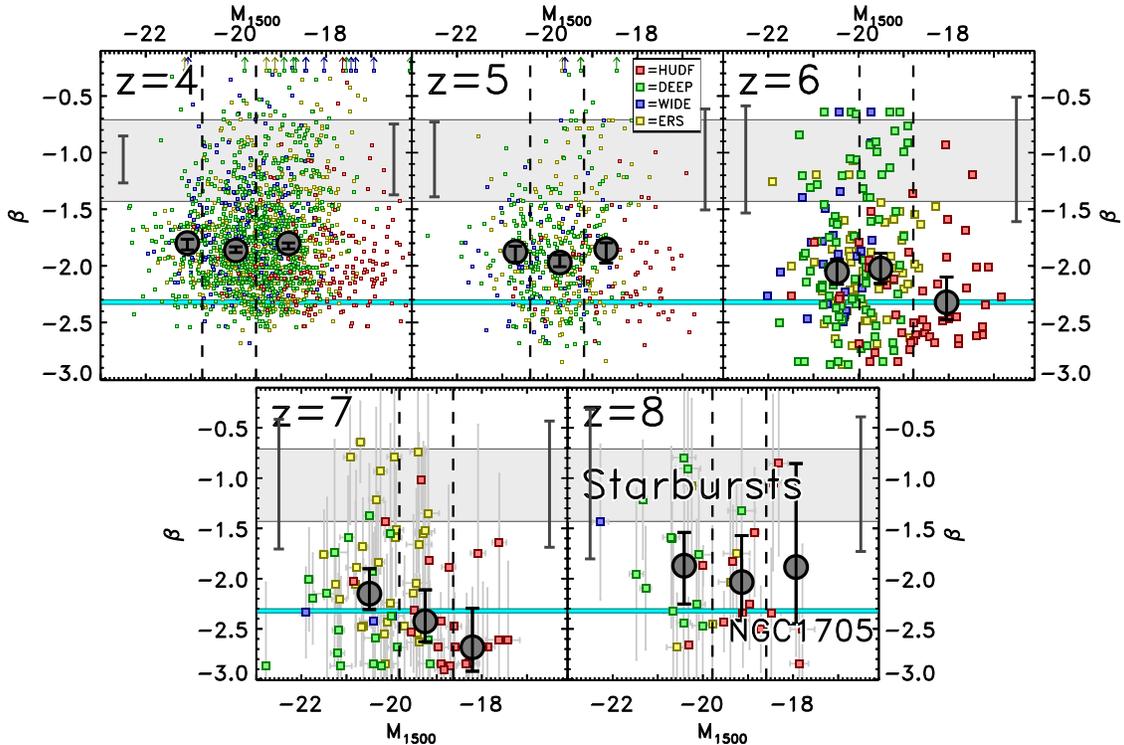}
\caption{The rest-frame UV spectral slope $\beta$ versus the derived absolute magnitude at rest-frame 1500 \AA, where the different panels display the results for the five redshift samples we studied. The small squares show the results for individual galaxies, where the color represents the field the galaxy was discovered in (Red = HUDF; Green = DEEP; Blue = WIDE; Yellow = ERS).  For clarity we only show the errors on the individual points in the two highest redshift bins, but we do plot a characteristic error at bright (M$_{1500} = -$21) and faint (M$_{1500} = -$19) mag towards the top of each panel (shifted towards the edge for clarity).  The gray circles show the median value of $\beta$ measured from three subsamples split by the dashed lines, which represent the evolving values of 0.25L$^{\ast}$ and 0.75L$^{\ast}$.  The uncertainties on these values are the errors on the median, derived via bootstrap Monte Carlo simulations.  The light gray shaded region denotes the range in $\beta$ spanned by typical local starbursts, while the cyan region denotes NGC1705, which is one of the bluest local galaxies, and is thought to contain little dust.}
\end{figure*} 

\subsubsection{Systematic Uncertainties}
The pure photometric uncertainties on our fluxes do not account for any other systematic uncertainties, such as zeropoint offsets, aperture corrections, and other uncertainties that scale multiplicatively with the flux.  These are typically accounted for by adding a small systematic error term when computing $\chi^2$; we have thus added a 5\% systematic error term when performing the SED fitting.  We note that while some of these effects may be random in nature, some may preferentially make fluxes in a given band brighter or fainter.  For the example of aperture corrections, this will not affect the measured value of $\beta$, as we apply a constant aperture correction to all bands for a given object (as we measure fluxes on PSF-matched images); thus the derived color will stay constant.  While any zeropoint offsets may have a different sign and amplitude, {\it HST} has a remarkable photometric stability, thus the zeropoints are accurate to $<$1\% \citep{kalirai09}, well within the additional 5\% error term we already include.

Nonetheless, it is interesting to examine our sample to see how such a systematic shift in fluxes could affect the resultant median value of beta.  We performed this test at $z =$ 4, as it is where the error on the median $\beta$ is the smallest (see \S 3.3), and hence where any systematic uncertainty could have the largest impact.  In this test, we systematically shifted the fluxes in a single band (running this test separately for both i$_{775}$ and J$_{125}$) up and down by 2\%, and then re-performed SED fitting to measure $\beta$.  We then compared the resultant values of $\beta$ to the true input value known from the simulation.  When shifting the i$_{775}$-band flux systematically for all objects up or down by 2\%, we found that the recovered $\beta$s differed from the true value by $\sim$1\%.  The difference was even smaller, at $\sim$0.3\% when making the shifts to the J$_{125}$-band.  Thus, in order to dominate over our measured uncertainty of $\sigma_{\beta}$ at $z =4$, any systematic error would have to be of the $\gtrsim$ 5\% level, much greater than any expected systematic uncertainty.

\subsection{Computing Median Values of $\beta$}
Figure 5 plots the calculated values of the UV spectral slope $\beta$ for each of our five galaxy samples from $z =$ 4 -- 8.  In these plots, we show the values of $\beta$ for local star-forming galaxies for reference.  The light gray region denotes the range of UV spectral slopes spanned by the templates of \citet{kinney96}, which represent typical local starbursts with a range of dust attenuation.  The cyan bar denotes NGC1705, which is one of the bluest local star-forming galaxies, and is thought to have sub-solar metallicity and be dust free \citep{calzetti94}.

As the uncertainty on the value of $\beta$ in an individual galaxy (particularly at fainter magnitudes and higher redshifts) is large, we compute median values of $\beta$ in three bins, split by 25\% and 75\% of the evolving value of the characteristic magnitude L$^{\ast}$, using the luminosity functions of \citet{bouwens07} for $z =$ 4, 5, and 6, and \citet{bouwens11} for $z =$ 7 and 8.  These median values are shown in Figure 5 by gray circles.  We note that our chosen value of 0.25L$^{\ast}$ for the faintest bin is similar to an apparent magnitude at $z =$ 7 of $\sim$ 28.3 AB mag, which is the apparent magnitude cut used in \citet{finkelstein10a} to split the bright from faint samples.

\begin{deluxetable*}{ccccccc}
\tabletypesize{\small}
\tablecaption{Median Values of the UV Spectral Slope $\beta$}
\tablewidth{0.9\textwidth}
\tablehead{
\colhead{Sample} & \colhead{Median $\beta$} & \colhead{Median $\beta$} & \colhead{Median $\beta$} & \colhead{Median $\beta$} & \colhead{$\beta$--M$_{1500}$$^{\dagger}$} & \colhead{Adopted$^{\ddagger}$}\\
\colhead{$ $} & \colhead{All Galaxies} & \colhead{L $>$  0.75L$^{\ast}$} & \colhead{0.25L$^{\ast}$ $<$ L $<$ 0.75L$^{\ast}$} & \colhead{L $<$ 0.25L$^{\ast}$} & \colhead{Slope} & \colhead{M$^{\ast}$}\\
}
\startdata
$z =$ 4&$-$1.82$^{+0.00}_{-0.04}$&$-$1.80$^{+0.03}_{-0.06}$&$-$1.86$^{+0.03}_{-0.02}$&$-$1.80$^{+0.01}_{-0.05}$&\phantom{$-$}0.01 $\pm$ 0.03&$-$21.06\\
$z =$ 5&$-$1.91$^{+0.02}_{-0.06}$&$-$1.88$^{+0.05}_{-0.09}$&$-$1.97$^{+0.07}_{-0.04}$&$-$1.85$^{+0.06}_{-0.12}$&\phantom{$-$}0.00 $\pm$ 0.06&$-$20.69\\
$z =$ 6&$-$2.07$^{+0.06}_{-0.10}$&$-$2.05$^{+0.11}_{-0.11}$&$-$2.02$^{+0.13}_{-0.13}$&$-$2.32$^{+0.22}_{-0.15}$&$-$0.10 $\pm$ 0.07&$-$20.29\\
$z =$ 7&$-$2.37$^{+0.26}_{-0.06}$&$-$2.15$^{+0.25}_{-0.16}$&$-$2.42$^{+0.31}_{-0.20}$&$-$2.68$^{+0.39}_{-0.24}$&$-$0.20 $\pm$ 0.11&$-$20.14\\
$z =$ 8&$-$1.95$^{+0.23}_{-0.27}$&$-$1.87$^{+0.33}_{-0.38}$&$-$2.03$^{+0.46}_{-0.38}$&$-$1.88$^{+1.03}_{-0.56}$&$-$0.03 $\pm$ 0.26&$-$20.10
\enddata
\tablecomments{$^{\dagger}$ We note that although a significant slope
  is not observed at $z =$ 4 and 5, if we restrict the faintest
  magnitude bin to only include galaxies from the HUDF dataset, we do
  recover a slope of $-$0.07 $\pm$ 0.01 at $z =$ 4, and $-$0.09 $\pm$
  0.03 at $z =$ 5.  $^{\ddagger}$ The reference for  our assumed  values of
  the characteristic magnitude M$^{*}$  are Bouwens et al.\ (2007) for
  $z =$ 4, 5 and 6, and Bouwens et al.\ (2011) for $z =$ 7 and 8.}
\end{deluxetable*}

To calculate the uncertainty on the median values of $\beta$, we run bootstrap Monte Carlo simulations, similar to those done in \citet{finkelstein10a}, which quantifies the uncertainty in how well we can measure the mean (and is not an estimate of the intrinsic scatter in the population).  In brief, we first account for Poisson noise (which is most relevant in the highest-redshift bins) by selecting $N^{\prime} = N + A \times \sqrt{N}$ galaxies in each bin through random sampling with replacement, where N is the number of galaxies in the bin, and A is a random number drawn from a normal distribution with a mean of zero and a standard deviation of unity.  We then account for the photometric noise by taking each of the galaxies in the modified sample, and obtaining a new estimate, $\beta^{\prime}$, for $\beta$, where $\beta^{\prime}$ = $\beta + B \times \sigma_{\beta}$, where B is a different random number with the same distribution as A and $\sigma_{\beta}$ is the photometric uncertainty in the value of $\beta$ for that galaxy.  The median value of $\beta^{\prime}$ is then computed from this modified sample.  The simulation is repeated 10$^{4}$ times.  In this manner, we build up a distribution of median $\beta$'s in each of the bins, and the 68\% confidence range is computed from these simulated $\beta$ values.  This process is done for all three bins (as well as for all galaxies together) in each of the five redshift samples.  We note that the Monte Carlo portion of these simulations are necessary to account for the intrinsic scatter in the rest-frame UV colors of galaxies, which contributes to the observed scatter in $\beta$, in addition to the photometric error.  The latter would be the only source of the observed scatter if one was was observing a population of galaxies with a uniform color; this is not the case at lower redshift, thus we do not expect it to be the case out to the highest redshifts.

We list the median values of $\beta$ in all three luminosity bins, as well for all galaxies in each redshift sample, in Table 4.  We summarize the evolution of the median $\beta$ of all galaxies in each sample with redshift in Figure 6.  

\subsection{Contamination}
Although the comparison with existing spectroscopy indicates contamination by lower-redshift galaxies in our sample is low, widespread spectroscopic coverage at the high-redshift end of our sample is not yet available.  One way we can investigate the possible effect of contamination on our results is to make our selection criteria more robust, and see how the results, for example the median $\beta$ at each redshift, vary.  We do this by varying the criterion of $\mathcal{P}_{primary}$ $\geq$ 0.7.  If we change this to a much more conservative value of $\geq$ 0.9, then we find that the median values of $\beta$ become $\sim$ 0.05 -- 0.15 bluer, as the galaxies which are removed tend to be redder.  

However, this does not mean that they are all contaminants.  Red galaxies naturally will have a $P(z)$ which is less robust.  A blue galaxy with a well defined break can only have one redshift solution, while a red galaxy, even with a well-defined break, will frequently have a secondary solution assuming that the detected break is the Balmer, rather than Lyman, break.  However, it is likely that our sample suffers from some level of contamination, and this exercise has shown that the change in median $\beta$ at each redshift is small relative to the uncertainties, and that the trend we find between $\beta$ with redshift does not significantly change.  We do note that this more conservative cut does remove 5/8 $z =$ 8 galaxies with $\beta >$ 1.5; thus if they are truly contaminants, the median $\beta$ at $z =$ 8 may be bluer than we measure.

\section{Discussion}

\subsection{Colors of Faint Galaxies at $z \sim$ 7}
We first examine the bluest data point, which is the median value of $\beta$ for the faint ($<$ 0.25L$^{\ast}$) galaxies at $z =$ 7, as this particular data point has received much attention in the literature.  \citet{bouwens10b} first published a value of the UV spectral slope for faint $z \sim$ 7 galaxies of $\beta = -$3 $\pm$ 0.2, discussing the possibility that these very blue galaxies could host metal-free star formation and/or a top-heavy initial mass function, as such exotic populations might be necessary to produce such blue values of $\beta$.  In \citet{finkelstein10a}, we examined the same dataset (the first year's release of the WFC3 data in the HUDF) and found a similar value of $\beta$, but with our error treatment above, we derived a larger uncertainty $\sigma_{\beta}$ = 0.5 \citep[see also][]{wilkins11}.  We thus concluded that evidence for exotic stellar populations was not substantiated within the uncertainties (it is worth noting that one can obtain a value of $\beta$ $\sim$ $-$3 even with ``normal'' stars, as a very young stellar population can create UV continua this blue, even when accounting for nebular continuum emission\footnote[3]{A stellar population with a Salpeter initial mass function (IMF), $Z$ = 0.02 $Z$\sol, a continuous star-formation history, zero dust and an age of 1 Myr has $\beta$ = $-$3.15, or $\beta$ = $-$2.90 when accounting for nebular continuum emission.  These values change only slightly when assuming Solar metallicity.}).  Also using these data, \citet{dunlop12} found via simulations that there is a bias towards artificially blue slopes for faint galaxies, from which they argued the very blue colors are overestimated.  Our simulations in \S 3.2 on the full-depth HUDF data do not show as strong a bias as that found by Dunlop et al., though we do find a bias of $\beta_{observed} < \beta_{intrinsic}$ by $\sim$ 0.25 for faint galaxies at $z =$ 7 and 8.  

\begin{figure*}[!ht]
\epsscale{0.8}
\plotone{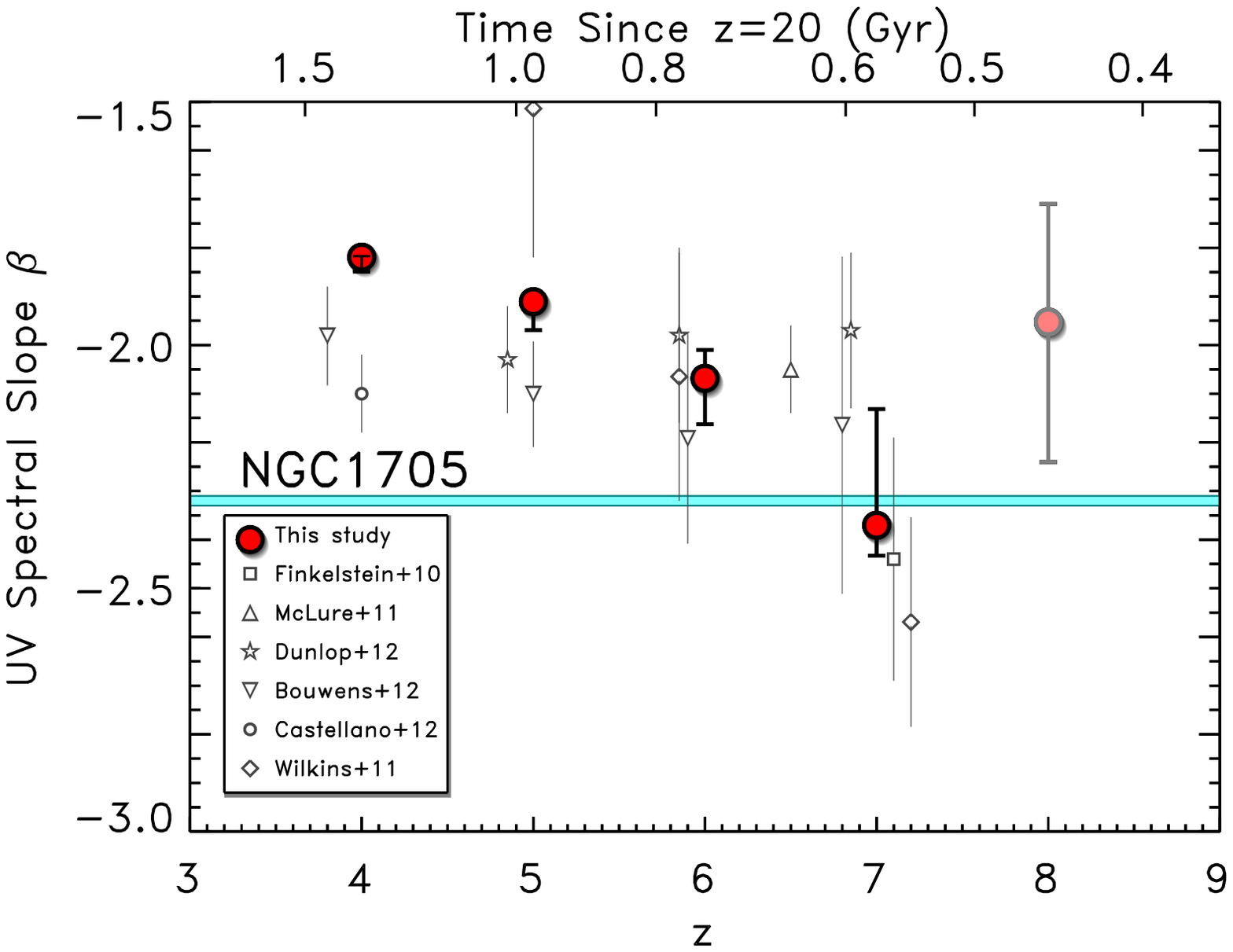}
\vspace{-5mm}
\caption{Evolution of the median value of the UV spectral slope $\beta$ with redshift for all galaxies in our sample (circles).  The median M$_{1500}$ for these points are: $-$19.4 ($z=$4), $-$19.7 ($z=$5), $-$19.9 ($z=$6), $-$19.9 ($z=$7) and $-$20.0 ($z=$8).  The smaller symbols represent recent results from \citet{finkelstein10a}, \citet{mclure11}, \citet{dunlop12}, \citet{bouwens12}, \citet{castellano12} and \citet{wilkins11}, each interpolated to represent the results at the median value of M$_{1500}$ for our sample (with the exception of the $z=$7 result of Finkelstein et al.\ 2010, which is shown at M$_{1500} = -$18.8).  We find that significant evolution in $\beta$ takes place from $z =$ 7 to 4, which is likely due to increased dust extinction.  We note that the point at $z =$ 8 is preliminary, due to the large scatter inherent in measuring $\beta$ for faint galaxies from a single color.  We have faded this point in the figure to caution the reader about these uncertainties.}
\end{figure*}

As shown in Figures 2 -- 4, our SED-fitting method has an advantage over the previously published methods which use only a single color.  Thus, with our new method and the improved dataset, it is interesting to see how the value of the faint bin at $z =$ 7 has changed.  As shown in Figure 5, we measure $\beta =$ $-$2.68$^{+0.39}_{-0.24}$ for faint galaxies at $z =$ 7.  This measured value is redder than that previously measured ($-$3 $\pm$ 0.5), yet we note that it is consistent with our previously published uncertainties.  The updated uncertainty is smaller due to a combination of an improved measurement method and the deeper data in the HUDF.  Although our sample is also larger due to the inclusion of the CANDELS and ERS fields, the faint bin is dominated by sources discovered in the HUDF.  Accounting for the selection bias discussed above (see Figure 4b), the value of $\beta$ may be even redder, at $\beta \sim -$2.4.  Thus, the value derived here is fully consistent with the expected UV slopes of ``normal'' stellar populations.  There is no evidence for extreme or exotic stellar populations.

\subsection{Evolution of $\beta$ with Redshift}
As shown in Figure 6, the median value of $\beta$ decreases significantly with redshift from $\beta$ $\sim$ $-$1.8 at $z =$ 4, to $\beta$ $\sim$ $-$2.4 at $z =$ 7.  We note that while the measured median $\beta$ at $z =$ 8 appears to be redder, the small number of $z =$ 8 galaxies combined with only a single detected color currently available for these candidate galaxies results in large uncertainties.  It seems unlikely that galaxies get bluer from $z =$ 4 -- 7, then redder again at $z =$ 8.  However, we note that if nebular continuum emission begins to play a dominant role at $z =$ 8, this could result in redder colors \citep{schaerer02}.  If future observations confirm this redder color, this could also signify an approach to the epoch of first galaxy formation, as an entire galaxy must be extremely young for nebular emission to dominate.  Focussing again on $z =$ 7, we find that galaxies in this early epoch have colors very similar to NGC1705, thus they are consistent with having little-to-no dust.  Our derived values of $\beta$ are broadly consistent with those from previous studies, which have found that galaxies at $z \sim$ 6 tend to be bluer than those at $z \sim$ 3 \citep[e.g.,][]{lehnert03, stanway05, bouwens06, hathi08, bouwens09, wilkins11, bouwens12, castellano12}.  \citet{dunlop12} recently studied the colors of robustly detected ($>$8$\sigma$) galaxies at $z >$ 5 finding that the galaxies at all redshifts are consistent with $\beta = -$2.05 $\pm$ 0.1 (shown as stars in Figure 6).  This is consistent within 1$\sigma$ of our measurements, though it is in tension with our $z =$ 7 measurement.  However, this difference is likely due to the excluded sources in the \citet{dunlop12} study; if we restrict ourselves to the brightest bin only, we would also find $\beta$ $\sim$ $-$2 (Figure 5 and Table 4).

There is a significant difference between results at $z=4$, where \citet{castellano12} have recently published an analysis on the colors of $z=4$ galaxies using a much more conservative selection, using the V$_{606}$ - H$_{160}$ color in place of V$_{606}$ - z$_{850}$.  From a sample of $\sim$ 170 galaxies which satisfy their selection in the HUDF and ERS fields, they compute an average $\beta$ of $-$2.1 $\pm$ 0.08.  To investigate whether our results would differ significantly, we applied the \citet{castellano12} color selection to our sample of $\sim$2000 $z =$ 4 galaxies, finding that $\sim$ 1000 galaxies satisfied the selection.  The median value of $\beta$ changed from $-$1.82 to $-$1.96; not as blue as found by \citet{castellano12}, but a significant change nonetheless.  However, even when using this more conservative sample selection criteria, there is still a substantial drop in $\beta$ from $z =$ 4 -- 7.

As a possible explanation for the reddening of galaxy colors with redshift, we investigate a simple model where we assume all galaxies in our sample formed at $z =$ 20, have 20\% of Solar metallicity, and are forming stars at a constant rate.  We then find how much dust extinction needs to be added to the population to create colors consistent with our results at $z =$ 4 and 7.  We find that at $z =$ 4, A$_{V}$ = 0.5 mag of extinction is required, while only A$_{V}$ $<$ 0.1 mag is required by $z =$ 7.  Thus this model is consistent with very little dust having been formed by $z =$ 7.  In reality, both the stellar population age and metallicity will be changing, but the rest-frame UV slope is fairly insensitive to metallicity \citep[c.f.][]{bouwens09}, and the rest-frame UV color is dominated by the most recently formed stars; thus if the SFH is constant or rising, age is not likely to significantly redden the slope (see also Figure 11 from \citet{bouwens12}).  We note also that a varying IMF can also affect the measured value of $\beta$, but thus far the observed values are consistent with a Salpeter IMF (see \S 4.1).  These derived extinction values are consistent with those obtained from the relation between $\beta$ and dust extinction from \citet{meurer99}, of A$_{V}$ = 0.33 mag at $z =$ 4, and A$_{V} =$ 0 at $z =$ 7.  Although this relation was derived at $z =$ 0, it has been shown to work at $z \sim$ 2 for most (i.e., not the youngest) galaxies \citep[e.g.][]{reddy06}.  If the galaxies at $z =$ 7 are uniformly young, then it might be that their infrared-excess is lower for a given value of $\beta$ than was assumed for the \citet{meurer99} relation.  In this case, the \citet{meurer99} relation would overestimate their extinction; thus little-to-no dust in typical $z =$ 7 galaxies in our sample appears likely.

If this change in color is dominated by increased extinction by dust, what mechanism of dust formation is responsible?  There are two primary sources for dust formation - supernovae (SNe) and asymptotic giant branch (AGB) stars.  Theoretical models have shown that a single SNe can produce $>$ 1 M\sol\ of dust \citep{todini01}, while observations have given mixed results, finding very small amounts of hot dust a few hundred days after the explosion, but larger amounts of cold dust many years after the explosion \citep[][and references therein]{gall11}.  Given that very dusty quasars have been seen at $z >$ 6 \citep[e.g.][]{wang08}, it is likely that SNe are an efficient producer of dust in the early universe.

However, the UV continua of typical $z >$ 6 galaxies in our sample appear to be relatively unattenuated.  The difference may be due to dark matter halo mass - the dusty quasars at high redshift inhabit the most massive halos and thus likely experience accelerated evolution, while even our ``bright'' subsample of galaxies have L$_{UV}$ $\lesssim$ L$^{\ast}$, and thus inhabit much lower mass halos.

In the local universe, AGB stars are the primary source of dust injection into the interstellar medium \citep[ISM; e.g.,][]{gehrz89}.  However, at high redshift, the universe is too young for low-mass stars to evolve to the AGB phase; thus AGB-created dust can only come from stars with main-sequence lifetimes less than the age of galaxy formation at a given redshift.  For example, at $z =$ 7, again assuming star-formation begins at $z =$ 20, only higher-mass stars with initial masses $>$ 3.2 M\sol\ will have evolved off the main sequence (the time elapsed from $z =$ 20 -- 7 is $\sim$ 600 Myr for our assumed cosmology).  AGB stars with initial masses of 2--4 M\sol\ are the most efficient dust producers, while M > 5 M\sol\ AGB stars produce little dust \citep[e.g.][]{gall11}\footnote[4]{We note that lower-metallicity stars will evolve to the AGB phase a few hundred Myr sooner, though it is also possible that galaxy formation begins a few hundred Myr later, at $z \sim$ 12 -- 15 (A. Karakas, private communication).}.

Thus our observation of an increasing dust content from $z =$ 7 $\rightarrow$ 4 is consistent with a scenario where, at least in these sub-L$^{\ast}$ galaxies, SNe at $z >$ 7 have not built up much dust.  As $\sim$ 3--5 M\sol\ stars enter their AGB phase at $z \sim$ 6--7, dust buildup then begins in earnest.  Then, in the epoch of $z =$ 6--4, when $\sim$ 1--3 M\sol\ stars enter the AGB phase, dust production ramps up, thus by $z =$ 4 a typical galaxy has a significant amount of dust.  We note that it may be that SNe at $z >$ 7 {\it are} significant dust producers, and that because these galaxies are low-mass much of the dust may be blown out of the galaxy.  We discuss this scenario further in \S 6.1.

We note that dust grain growth in the ISM is an alternate, non-stellar, method for dust formation.  In this scenario, dust accumulates in the ISM on seeds formed by previous supernovae \citep[e.g.,][]{draine79, dwek80, draine09}.  The timescale for this process is only a few $\times$ 10$^{7}$ yr, thus this process can account for dust formation in the epoch 4 $< z <$ 8.  \citet{michalowski10} find that the combination of dust from SNe and AGB stars are not enough to explain the amount of dust detected in $z >$ 5 quasars, and thus conclude that grain growth in the ISM may be a contributing factor.  If $z >$ 7 SNe are providing dust seeds, then grain growth could be an important dust-growth mechanism throughout the epoch of our observations.
 
Although the timescales work for the color evolution to be due to AGB-formed dust, this conclusion is dependent on the formation redshift of these galaxies.  If they do form the first generation of stars at $z \sim$ 15 -- 20, then they should remain quasi-uniformly blue until significant dust is generated.  In the AGB-dominant scenario, this is at $z \lesssim$ 7.  Thus, better statistics at $z =$ 8 can help resolve this issue.  If the colors of galaxies at $z =$ 8 are similar to those at $z =$ 7, then it implies that not much dust is forming in that epoch, thus implying that SNe at high redshift are not efficient dust producers.  On the other hand, if the colors of galaxies at $z =$ 8 are bluer than those at $z =$ 7, the gradual reddening of color from $4 < z < 8$ might better be explained by SNe, with AGB stars playing a smaller role.  This will be possible in the near future with upcoming data in the HUDF (PI Ellis) which are expected to greatly increase the robustness of $\beta$ measurements at $z =$ 8.  We note that the current data at $z =$ 8 (although tenuous) are more consistent with no evolution in $\beta$ from $z =$ 7 -- 8 than with continued evolution toward bluer colors.

\section{Dependence of $\beta$ on Magnitude and Stellar Mass}

\subsection{Correlation of $\beta$ with M$_{1500}$}
As is apparent in Figure 5, we do not observe a strong relation between $\beta$ and the UV absolute magnitude at $z =$ 4 and 5.  However, at $z =$ 6 and 7, it does appear as if fainter galaxies have bluer values of $\beta$.  To quantify this, we fit a first-order polynomial through the median data points at each redshift to measure the slope and its associated uncertainty.  These values are listed in Table 4.  We find no significant ($\lesssim$ 2$\sigma$) slope between the galaxy colors and their absolute magnitudes at all redshifts.  This is peculiar, as a color-magnitude relation has previously been observed.  For example, \citet{papovich01}  and \citet{papovich04} observed a strong color-magnitude relation at $z \sim $3--4.  However, the color used was between the rest-frame UV and rest-frame optical, spanning the 4000 \AA\ break.  This color is strongly affected by age; we solely use colors blueward of the 4000 \AA\ break at the redshifts probed by our study, thus we are much more sensitive to evolution in the dust extinction than in the population age.  

\citet{labbe03} studied the rest-frame UV colors of galaxies out to $z \sim$ 3, finding a color-magnitude relation at all redshifts using a rest-UV single color.  \citet{bouwens09} performed a similar analysis, finding a similar relation at $z \sim$ 2.5 and 4, as did \citet{wilkins11}, who used single colors to study the dust extinction at $z =$ 5 -- 7.  On the other hand, \citet{reddy08} found no significant change in the dust reddening over the magnitude range they probed at $z \sim$ 2--3, (though their sample was on average brighter than ours; i.e., closer to L$^{\ast}$), and \citet{dunlop12} also did not detect a trend between $\beta$ and the UV magnitude, though they restricted their sample to galaxies which had at least one 8$\sigma$ detection.  Using a power-law fit to the data to measure $\beta$, \citet{castellano12} only find modest evidence for a trend between $\beta$ and M$_{1600}$; with $\Delta \beta$ $\simeq$ 0.2 across the magnitude range probed by their study (M$_{1600} = -$18.5 to $-$21).  Recently, \citet{lee11} found evidence for redder UV spectral slopes for luminous ($\sim$ 5L$^{\ast})$ galaxies when compared to $\sim$ L$^{\ast}$ galaxies at $z \sim$ 4 using two colors; our somewhat limited dynamical range in luminosity prohibits us from observing such a trend.  

Using a similar dataset to our own, \citet{bouwens12} recently published a study of galaxy rest-UV colors at $z =$ 4 -- 7,  computing $\beta$ by fitting a power-law to the observed photometry (and correcting for any observational biases via simulations).  From our simulations discussed in \S 3.2, it is unlikely that the different method of measuring $\beta$ would affect the relation between $\beta$ and the M$_{UV}$, yet they find strong evidence for such a correlation at all redshifts while we do not.  However, the difference in the way we measure M$_{UV}$ may have an effect.  As we integrate the best-fit model through a 100 \AA\ wide bandpass centered at 1500 \AA, we are always probing rest-frame 1500 \AA.  The geometric mean between the filters used in the $\beta$ derivation by \citet{bouwens12} probes different parts of the rest-frame UV depending on the redshift.  For $z =$ 7, this probes closer to 1500 \AA\, while at $z =$ 4, it probes wavelengths redder than 2000 \AA, far from where our measurement probed.   To see if this has an effect, we re-created Figure 5 measuring M$_{UV}$ in the same way as \citet{bouwens12}, finding that defining the absolute magnitude in this way allows us to recover a $>$3$\sigma$ significance measurement of the $\beta$--M$_{UV}$ slope at $z =$ 4 (of $-$0.04 $\pm$ 0.01; significant, but not as strong as the slope found by \citet{bouwens12}).  This is understandable, as at $z =$ 4, our defined absolute magnitude sits at one end of the wavelength baseline used for $\beta$ (close to i$_{775}$), which may wash out any dependance between $\beta$ and the absolute magnitude.  For example, if a given galaxy is brighter in the rest-frame UV, this corresponds to a brighter i$_{775}$-band flux, which, as the one of the bluest filters, would result in a bluer spectral slope, washing out a correlation of redder colors with brighter magnitudes.  At $z =$ 7 this is not the case (i.e., M$_{1500}$ is measured in the J$_{125}$-band, which is in the middle of the three filters used to derive $\beta$), and correspondingly we do see an apparent dependence between $\beta$ and M$_{1500}$, though it is not significant.  To further investigate this, we repeated our analysis, defining M$_{UV}$ to be at 2000 \AA\ rather than at 1500 \AA.  This is closer to the wavelength for M$_{UV}$ at $z =$ 4 defined in \citet{bouwens12}.  We find that this can result in a marginally significant slope at $z =$ 4, of $-$0.04 $\pm$ 0.02.

Thus, the method used to define the rest-frame UV absolute magnitude can affect whether a dependance between $\beta$ and M$_{UV}$ is recovered.  Additionally, although the rest-frame wavelengths probed in these studies are similar to ours, the methods of deriving the colors are different for the majority of them, as we use multiple rest-UV colors in our SED-fitting method.  We tested how the measured value of $\beta$ depends on these different methods.  We recreated our $\beta$-magnitude plots (i.e.\ Figure 5), deriving $\beta$ from a single color, via SED-fitting, and also through a power-law fit to the photometric data.  We made these plots using both our own photo-z selected sample, but also using only those galaxies that satisfied standard LBG color selection as well (using the selection criteria from \citet{bouwens12}).  We found that whether one uses a photo-z or LBG method for the selection does not affect the color-magnitude dependance observed.  We also found that we can detect a color-magnitude dependance only when using a single color.  Specifically, when using a single color to measure $\beta$ at $z =$ 4, we find a slope of $-$0.11 $\pm$ 0.05, or a 2$\sigma$ significance.  Thus, it would appear as if the method of measuring $\beta$ can affect whether a color-magnitude dependence is observed.  This may be due in part to the bias at faint magnitudes discussed by \citet{dunlop12}.  If one only uses a single color to derive $\beta$, and that color is also included in the selection, then a bias towards bluer colors when the signal-to-noise is low can be introduced.

We also investigated whether the dynamic range in M$_{1500}$ or our choice of magnitude binning is affecting our results.  We did this by only including galaxies from the HUDF in the faintest bin; this changes things very little at $z \geq$ 6, but it does effectively move the faintest bin to $\sim$ 1 magnitude fainter at $z =$ 4 and 5.  We find that this results in a significantly (7$\sigma$ and 3$\sigma$, respectively) detected slope of $-$0.07 $\pm$ 0.01 at $z =$ 4, and $-$0.09 $\pm$ 0.03 at $z =$ 5. The emergence of a trend when restricting the faintest bin in this way implies that if the correlation between $\beta$ and M$_{1500}$ is physical, it may be flatter at bright magnitudes.  This can be further studied in the future with the inclusion of a larger sample of brighter galaxies from the other CANDELS fields, and more faint galaxies from the HUDF parallel fields \citep{oesch07,bouwens11}.

To investigate this further, we examined the values of $\beta$ derived in the faint bin at $z =$ 4 when including only HUDF galaxies, and also when only including CANDELS+ERS galaxies, finding a difference of 0.26 $\pm$ 0.10, or 2.6$\sigma$.  However, we acknowledge that the $z =$ 4 bin is unique, in that the lack of U-band data may result in a somewhat higher contamination in this bin, as galaxies at the low-redshift end of this bin will not have the Lyman limit completely blueward of B$_{435}$.  Some evidence for this was discussed in \S 2.4, where we note that the 12 contaminant galaxies identified via the discussed spectroscopic coverage all had photometric redshifts in the $z =$ 4 bin.  We note that we do not expect this contamination to be large, as the 12 removed galaxies were a small fraction of the total number of galaxies in our $z =$ 4 sample with spectroscopic coverage.  One caveat is that the majority of galaxies targetted for spectroscopy are relatively bright, thus we do not have an emprical constraint on the contamination of the fainter galaxies in our sample, where the largest discrepancy in $\beta$ between the fields is found.

We conclude with noting that there are a variety of systematic and physical reasons why a trend may or may not be observed, with a primary factor being the definition of the UV absolute magnitude.  While we elect to use a constant rest-frame wavelength to better track evolution, this may lessen our ability to track a correlation between $\beta$ and M$_{UV}$.

\subsection{$\beta$--Stellar Mass Trend}
The fact that the detectability of a dependence between $\beta$ and the UV absolute magnitude can depend on the definition of the latter implies that this may not be the most robust method to trace the dependence of $\beta$, as one is comparing two correlated quantities.  We thus investigate whether comparing $\beta$ to the stellar mass results in a more robust trend.  While the stellar mass is dependent on the absolute brightness of a galaxy, it is not directly dependent on the galaxy color, while the UV absolute magnitude is dependent when one averages two different filters to derive it.  As we have used the SED-fitting method to derive the values of $\beta$, we have also derived the stellar masses for all the galaxies in our sample.  In Figure 7, we again plot each of the galaxies in our sample similar to Figure 5, only now we plot the stellar mass on the horizontal axis.  Average values of $\beta$ were computed using similar bootstrap Monte Carlo simulations as before; here we split objects into three bins; $7 < $log M/M\sol$ < 8$, $8 < $log M/M\sol$ < 9$ and $9 < $log M/M\sol$ < 10$.  The median values of $\beta$ in these bins are given in Table 5.

\begin{figure*}[!ht]
\epsscale{0.9}
\plotone{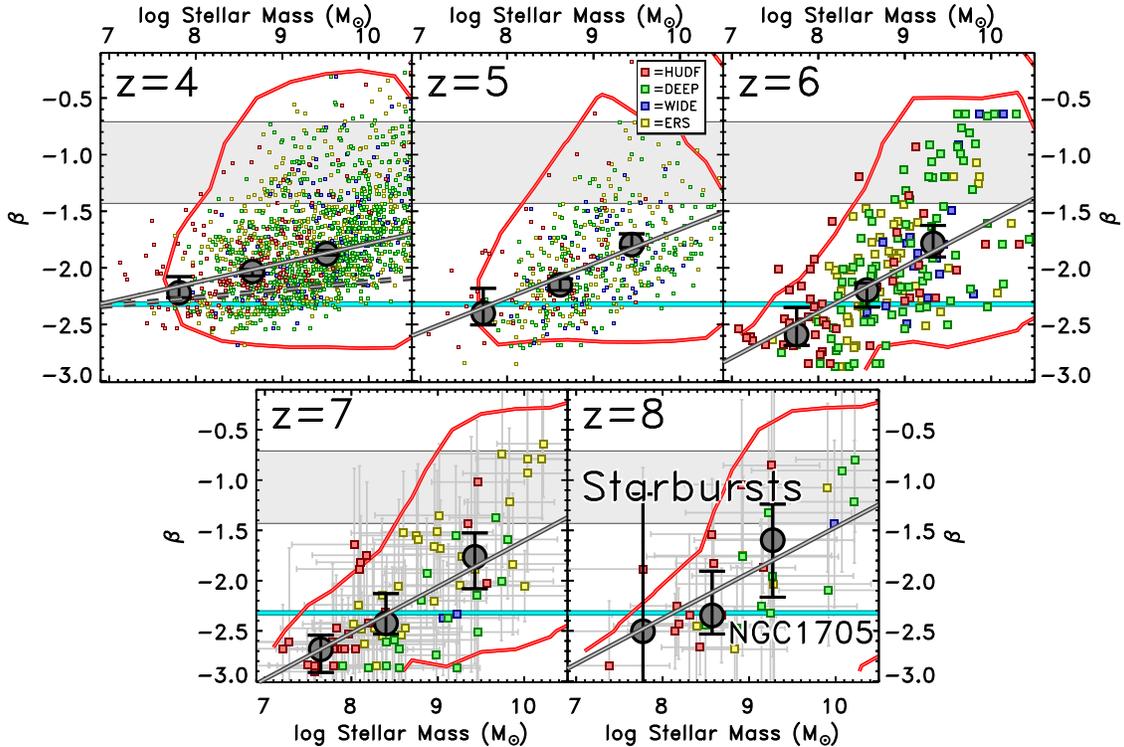}
\caption{Similar to Figure 5, except using stellar mass on the horizontal axis.  We find a significant correlation between $\beta$ and the stellar mass at all redshifts.  The solid gray lines show a linear fit through the median $\beta$ points.  The red curve shows the 20\% completeness limit for the HUDF galaxies, derived from our simulations (these contours will be slightly contracted for the other fields; one can estimate where they would lie by looking at the relative distribution between HUDF galaxies and those in the other fields).  Taking the completeness into account, it appears as if the correlation seen at $z =$ 4 -- 5 is significant, as we would have detected faint, red galaxies if they were there.  However, this is not the case at $z =$ 6 -- 8, as the completeness contour closely traces our observed trend, implying incompleteness is dominating this effect in the higher-redshift bins.  The dashed line in the $z =$ 4 panel shows our estimate of the contribution of the intrinsic correlation of $\beta$ with the mass-to-light ratio to the overall trend.}
\end{figure*}

\begin{deluxetable*}{cccccc}
\tabletypesize{\small}
\tablecaption{Values of the UV Spectral Slope $\beta$ Split by Stellar Mass}
\tablewidth{0pt}
\tablehead{
\colhead{Sample} & \colhead{Median $\beta$} & \colhead{Median $\beta$} & \colhead{Median $\beta$} & \colhead{\# Galaxies/Bin} & \colhead{$\beta$--Stellar Mass}\\
\colhead{$ $} & \colhead{log M/M\sol\ = 7 -- 8} & \colhead{log
  M/M\sol\ = 8 -- 9} & \colhead{log M/M\sol\ = 9 -- 10} & \colhead{Low/Middle/High Mass} & \colhead{Slope}\\
}
\startdata
$z =$ 4&$-$2.22$^{+0.13}_{-0.05}$&$-$2.03$^{+0.02}_{-0.04}$&$-$1.88$^{+0.02}_{-0.02}$&52/510/927&0.17 $\pm$ 0.03\\
$z =$ 5&$-$2.40$^{+0.22}_{-0.10}$&$-$2.15$^{+0.06}_{-0.04}$&$-$1.79$^{+0.09}_{-0.03}$&24/227/223&0.30 $\pm$ 0.06\\
$z =$ 6&$-$2.59$^{+0.23}_{-0.10}$&$-$2.20$^{+0.05}_{-0.16}$&$-$1.78$^{+0.19}_{-0.10}$&29/104/76&0.40 $\pm$ 0.10\\
$z =$ 7&$-$2.68$^{+0.15}_{-0.24}$&$-$2.42$^{+0.31}_{-0.11}$&$-$1.76$^{+0.23}_{-0.33}$&17/36/22&0.46 $\pm$ 0.10\\
$z =$ 8&$-$2.50$^{+1.26}_{-0.43}$&$-$2.35$^{+0.46}_{-0.16}$&$-$1.60$^{+0.32}_{-0.54}$&3/15/12&0.45 $\pm$ 0.37
\enddata
\end{deluxetable*}

We find a $>$ 2$\sigma$ significance dependance between $\beta$ and the stellar mass at all redshifts (except at $z =$ 8), in that more massive galaxies are redder.  However, we caution that completeness may play a strong role here, in that low-mass blue galaxies are easier to detect in our dataset than low-mass red galaxies.  To examine if incompleteness is dominating our trend, we computed completeness levels from our simulations in \S 3.2 as the number of recovered galaxies divided by the number of input galaxies in bins of stellar mass and $\beta$.  We derived the 20\% completeness level, and we plot this in Figure 7 as the red curve.  We note that while these simulations were only performed on the HUDF images, one can estimate where similar contours will lie for the other regions by the distribution of data points for objects from the different fields.

At $z =$ 4, it seems our observed correlation between $\beta$ and the stellar mass is real, as we would have detected a significant population of red ($-$1.5 $< \beta <$ $-$1.0), low-mass (8.5 $<$ log M/M\sol\ $<$ 9.5) galaxies had they existed.  The same is true at $z =$ 5.  However, our completeness begins to change at $z \geq$ 6, in that our ability to detect low-mass red galaxies is reduced due to the shortened wavelength baseline which we are probing.  Inspecting the completeness contours at $z =$ 6, 7 and 8, we find that the shape of the contour is similar to the shape of the derived $\beta$-stellar mass slope, implying that at the highest redshifts, incompleteness may be dominating the observed trend.  However, we note that as there are relatively few red, low-mass galaxies at $z =$ 4, its unlikely that a large population exists at $z =$ 7.  Additionally, in the $z =$ 7 panel of Figure 7, galaxies do not tend to pile up against the completeness contour, further implying that low-mass red galaxies are not common at very high redshift.

\begin{figure}
\epsscale{1.2}
\hspace{-5mm}
\plotone{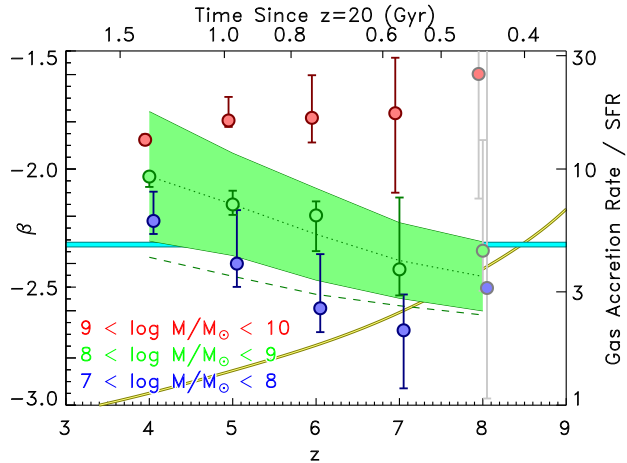}
\vspace{-2mm}
\caption{Similar to Figure 6, except here we split our sample into three bins by stellar mass.  The red circles denote the median $\beta$ as a function of redshift for galaxies with 9 $<$ log M/M\sol\ $<$ 10, green circles for 8 $<$ log M/M\sol\ $<$ 9, and blue circles for 7 $<$ log M/M\sol\ $<$ 8.  The green shaded region highlights the range in $\beta$s from the models of \citet{finlator11} for the middle mass range.  The dotted line shows the median value of $\beta$ from these models, while the dashed line shows the median $\beta$ from the same models with no dust extinction.  The cyan bar denotes the value of $\beta$ for NGC1705.  This plot indicates that the trend observed in Figure 6 is largely drived by the lower two mass bins.  The massive galaxies have a roughly constant $\beta$ at each redshift, implying that they are able to retain the dust formed in earlier epochs via SNe, while the lower-mass galaxies were not.  The yellow curve (and the right-hand vertical axis) shows the ratio between the gas accretion rate and the star-formation rate using the predictions from \citet[][see \S 6.3]{papovich11}.  Values greater than unity imply that galaxies at that redshift are accreting more gas than they are converting into stars, which may hinder the escape of \lya\ photons.}
\end{figure}

One caveat here is that we are only using photometry that covers the rest-frame UV, while photometry covering redder wavelengths results in more robust estimates of the stellar mass (by measuring the light from lower-mass stars).  One may thus worry that any derived correlation is due to an intrinsic correlation within the models between $\beta$ and the mass-to-light ratio.  To investigate this, we examined our suite of stellar population models, investigating the range of mass-to-light ratios at a given value of $\beta$.  Investigating $-$1.5 $> \beta > -$2.5, which is the observed range of $\beta$ within our sample, we find that at any given value of $\beta$, there is more than an order of magnitude spread in the mass-to-light ratios.  This is easy to understand, as we are allowing a wide range of star-formation histories and stellar population ages, along with metallicities and dust attenuations, thus there is a large combination of model parameters which can yield a given $\beta$, each with a different mass-to-light ratio.  \citet{mclure11} found a similar result, finding their allowed models spanned a mass-to-light range of $>$ 50 when looking at a given UV luminosity.  Additionally, the mass-to-light ratio distributions for models with $\beta \sim -$2.2 and those with $\beta \sim -$1.8 have a $>$50\% overlap.  For the non-overlapping models, the bluer tend to have lower mass-to-light ratios.  However, the difference results in the redder galaxies having a mass $\sim$3$\times$ greater than the bluer galaxies; this is less than the factor of 10 or higher which we show in Figure 8.

Even though there is such a large spread in M/L, we investigate how much of an impact it may have on our derived relation.  At $z =$ 4, where we are the most complete, we first took the SED-fitting derived masses, and computed the median M/L ratio.  We then applied this median M/L to the UV luminosity for each galaxy to rederive a stellar mass, which is now independent of the individual SED-fitting results.  This should remove any intrinsic variation of the M/L with the models.  We then measured the resultant slope between $\beta$ and the stellar mass in the same way as we did on our sample, and found a slope of 0.10 $\pm$ 0.07.  The difference between our observed slope and this value is thus an estimate of the impact of the intrinsic M/L ratios within the models, which is thus a slope of 0.07.  This is indicated in Figure 7 by a dashed line.  This is $\sim$40\% of our observed trend; thus it appears as if our observed trend is due to both an intrinsic dependance between $\beta$ and the M/L ratios, as well as a dependence between $\beta$ and the luminosity.  In any case, these trends have a physical cause; a very blue galaxy will be dominated by a young population of massive stars, which have low mass-to-light ratios.

The issue of correlations between $\beta$ and the UV M/L ratio could potentially be alleviated by including photometry from the {\it Spitzer Space Telescope} Infrared Array Camera \citep[IRAC][]{fazio04} when measuring the stellar mass.  While measuring IRAC fluxes for all of the galaxies in our sample is beyond the scope of this paper, we have measured de-blended IRAC fluxes for our sample of galaxies in the HUDF using TFIT \citep{laidler07}.  To see whether the inclusion of IRAC fluxes results in a significant change in the stellar masses, and thus our observed $\beta$--mass correlation, we examined the $\beta$--mass slope for HUDF galaxies only, both with and without inclusion of the IRAC fluxes.  We found that the derived slopes were consistent within $1\sigma$, thus we conclude that the lack of rest-frame optical photometry is not driving the observed correlations.

For an additional test as to the validity of our derived masses, we use the simulations of \citet{finlator11}, described in \S 6.2 below.  These simulations include realistic star-formation histories which yield reasonable agreement with the observed rest-frame UV luminosity function and the growth rate of the stellar-mass density at $z >$ 6, hence they represent a physically-motivated test of the robustness of our stellar mass estimates.  The star-formation histories in these simulations are smoothly-rising, which is reasonable given the tightness of the observed star-formation-rate -- stellar mass relation \citep{labbe10b}.  We have performed SED fitting using the simulated photometry of these galaxies, ``observing'' the galaxies, such that their photometric data points are varied by an amount proportional to the photometric error in a given band (assuming the image depths in the HUDF).  We then compare the best-fit stellar masses using the same {\it HST} ACS and WFC3 bandpasses as used on our sample to the ``known'' input value of the stellar mass.  We find that we can recover the stellar mass very well at log M/M\sol\ $>$ 9, while at log M/M\sol\ $<$ 8 we begin to do much worse.  For example, at $z =$ 4, the 1$\sigma$ spread in recovered stellar mass at an input mass of log M/M\sol\ $=$ 9 is $\sim$ 0.25 dex, increasing to $\sim$ 0.5 dex at log M/M\sol\ $=$ 8.  At higher redshift, the lack of rest-frame optical is less detrimental, as the age of the Universe begins to become a strong constraint on the amount of mass in old stars \citep{finkelstein10a}.  Thus, except for the lowest masses in our sample, our SED fits with only the rest-frame UV fluxes do a reasonable job of recovering the {\it average} stellar masses of galaxies in our sample, and any uncertainty due to the lack of rest-frame optical observations should not affect the median-$\beta$ trends we derive.

\section{Implications}

\subsection{Dust Formation}
As we have now shown that $\beta$ exhibits a robust dependance on stellar mass, it is prudent to revisit our evaluation of the evolution of $\beta$ with redshift.  In Figure 8 we show again $\beta$ versus redshift (similar to Figure 6), only now we split our sample into the same three mass bins, where blue, green and red denote the least massive, middle-mass, and most-massive galaxies (split again at log M/M\sol\ = 8 and 9).  While the redder $\beta$ with increasing mass at a given redshift is easily visible, what is more interesting here is the evolution in $\beta$ in galaxies with a constant mass.  The two lowest mass bins both show a trend qualitatively similar to Figure 6, in the galaxies at log M/M\sol\ $<$ 9 become increasingly bluer with increasing redshift.  We note again that for the lowest mass galaxies at $z \geq$ 6, it is possible that we are missing a population of low-mass red galaxies, but the lack of a dominant red, low-mass population at $z =$ 4 -- 5 indicates that such a population is unlikely to exist at $z \geq$ 7.

The most massive galaxies at any given redshift show a very different trend, in that they appear somewhat red ($\beta \sim -$ 1.7 -- 1.9) at all redshifts.  Thus our inference that perhaps SNe are not forming dust in the galaxies in our sample cannot be true, as massive galaxies at $z =$ 7 (and possibly $z =$ 8) already appear to be somewhat dusty.  Since at $z =$ 7 the universe is likely too young for low-mass stars to have reached their AGB phase, the most likely explanation for this red color is dust created in SNe.  If dust is created in SNe in the most massive galaxies, then it must be created in the lower-mass galaxies as well.  Yet, the lower-mass galaxies still appear to be dust-free (at $z =$ 7 -- 8 for the 8 $<$ log M/M\sol\ $<$ 9 sample, and at $z \geq$ 5 for the 7 $<$ log M/M\sol\ $<$ 8 sample).  One plausible explanation is that feedback from the SNe explosions is driving the dust out of the low-mass galaxies, as their lower gravitational potentials make it harder for them to retain gas and dust in their ISM against feedback.  This is testable as this same mechanism could cause the metallicities of low-mass galaxies to evolve more strongly, thus the correlation between $\beta$ and stellar mass might indicate the slope of the mass-metallicity relation at these high redshifts, which will be testable in the future with the {\it James Webb Space Telescope} ({\it JWST}).

What does this mean for the lowest-mass galaxies?  If SNe were the only dust production mechanism, then one might expect low-mass galaxies to remain somewhat dust-free.  However, we see that in the two lowest mass bins galaxies do get redder with time.  While the lowest-mass bin is the most susceptible to incompleteness, the 8 $<$ log M/M\sol\ $<$ 9 bin shows a very similar trend to what we saw in Figure 6 for all the galaxies in the sample.  These results are thus still consistent with the idea that AGB stars are dominating the dust creation in lower mass galaxies (or, at least dominating the dust that is retained), as it is only after $z =$ 7 that they begin to show evidence for dust extinction.  The red, massive galaxies at $z =$ 7 show that if SN-created dust were dominating the extinction in the lower mass galaxies, we would see evidence for this at $z =$ 7.  While dust created in AGB stars would of course be subject to the same feedback effects as the SN-created dust, this dust comes $>$ 500 Myr after the onset of the star-formation episode, thus any massive stars which formed near the low-mass (now AGB) stars will have exploded far in the past, while any newly formed massive stars need not be in close proximity.

\subsection{Comparison with Simulations}
In Figure 8 we also show the predicted evolution of $\beta$ from the simulations of \citet{finlator11}.  These simulations were run using a custom version of the cosmological hydrodynamic code {\sc gadget-2} that includes recipes for cooling, star formation, and momentum-conserving outflows.  The simulated stellar continua are derived by integrating over the discrete stellar populations in each galaxy using the Bruzual \& Charlot (2003) population synthesis models.  Nebular line emission is treated following \citet{schaerer10}, and dust is added via the foreground screen model of \citet{calzetti00} with the normalization E(B-V) tied to the galaxy metallicity following a local calibration \citep{finlator06}.  The resulting simulations reproduce reasonably well the slope, evolution, and normalization of the observed galaxy mass-metallicity relation at $z \lesssim$ 3 \citep{finlator08, dave11} as well as the observed specific star formation rates and luminosity functions of Lyman-break galaxies at $z \geq$ 6 \citep{dave06, finlator11}.  The good agreement that these models demonstrate with existing constraints indicates that we may use them to understand the possible implications of our new observations.

In Figure 8 we show the values of $\beta$ and stellar mass from the simulations; to be sure there was no bias we also measured these parameters from the simulated galaxy photometry in the same way as for our observed objects.  We found excellent agreement between our SED-fitting derived values of $\beta$ and stellar mass and the input values from the simulated galaxies.  These simulations have relatively few galaxies in our massive bin, while our observed galaxies are more rare in the lowest-mass bin, so we compare to the simulations in the middle bin of 8 $<$ log M/M\sol\ $<$ 9, where both the simulations and observations have a large number of galaxies.  The simulations are denoted by the green shaded region, which denotes the 68\% spread in $\beta$ values at a given redshift.  We find good agreement between our observations and these simulations over the entire redshift range probed.  We note that the simulations of Dayal and Ferrara (2012) also predict a dependance of $\beta$ on the stellar mass, albeit at a lower level ($\Delta \beta \sim$ 0.2 at $z =$ 7 from log M $=$ 7.5 $\rightarrow$ 9.5).

This agreement has a number of implications.  First, the simulations allow us to examine the galaxies with and without dust, which in turn allows us to examine how much of the trend of increasing $\beta$ with redshift is driven by an increase in dust.  We find that the fully processed simulated galaxies (i.e., including dust extinction) have a $\beta$ that rises $\Delta\beta$ $\sim$ 0.5 from $z =$ 8 -- 4 (dotted green line in Figure 8), while the underlying stellar spectrum rises by $\Delta\beta$ $\sim$ 0.2 during that same time (dashed green line in Figure 8).  Thus, while increasing age and metallicity is contributing to the reddening of the color, we confirm that dust extinction is the dominant factor in the model galaxies.

As mentioned above, these simulations assume a constant value of extinction to metallicity, using results from local galaxies in the Sloan Digital Sky Survey \citep[see][]{finlator06}.  Since the evolution of $\beta$ at a common stellar mass within these simulations appear to agree with our observations, and $\beta$ is dominated by dust, this then implies that the way in which galaxies suffer extinction does not change much over $\sim$ 13 Gyr of cosmic time.

Additionally, this consistency of $\beta$ between the simulations and our observed galaxies imply that our observed galaxies may have similar metallicities to those in the simulations at a given redshift and stellar mass.  We computed the average and standard deviation of the simulated galaxy mass-weighted stellar metallicities at each redshift for galaxies with 8 $<$ log M/M\sol\ $<$ 9.  We find that the typical galaxy in this mass range likely has a stellar metallicity of $Z =$ 0.22 $\pm$ 0.06 $Z$\sol\ at $z =$ 4, dropping to $Z$ $\sim$ 0.13 $\pm$ 0.04 $Z$\sol\ by $z =$ 7.  These metallicities are low, but are well above some of the most metal-poor local galaxies, adding further evidence that these galaxies do not host Population III stars.

Finally, these models include the effects of outflows, which reproduces the observed mass-metallicity relation at lower-redshifts \citep[e.g.][]{tremonti04, erb06} by removing more gas (and metals) in lower-mass galaxies \citep{finlator11, dave11}.  This agrees well with our observed dependance of $\beta$ with mass, which at its heart is a mass-dust relation, and thus implies that the mechanisms which drive the $z <$ 3 mass-metallicity relations exist out to the highest redshifts we can observe.  In fact, our derived $\beta$--mass slope at $z =$ 4 (where we are the most complete) of 0.17 $\pm$ 0.02 is consistent with the value of 0.15 we measure from the \citet{finlator11} models, further implying that the slope we observe is in fact a manifestation of the mass-metallicity relation (see also e.g., Bouwens et al.\ 2009; Bouwens et al.\ 2011 for similar evidence for a luminosity-metallicity relation at $z \geq$ 4).

This work motivates a more detailed theoretical inquiry into the detailed dependence of gas fraction, UV continuum 
color, and metallicity on stellar mass as well as the evolution of these trends.

\subsection{Probing Reionization}
Finally, we come back to the increasingly blue colors of our overall sample of candidate galaxies with increasing redshift.  The evidence points to these galaxies becoming progressively more deficient in dust (and possibly metals) out to $z =$ 7.  Thus one may expect that the incidence of strong Ly$\alpha$ emission from these galaxies should be increasing as well.  In fact, this is seen out to $z =$ 6 \citep{stark10, stark11}.  However, recent work has shown that the incidence of detectable Ly$\alpha$ emission in galaxies decreases strongly from $z =$ 6 to 7, with the inference being that the volume neutral fraction of hydrogen in the intergalactic medium (IGM) rises up to $\sim$ 50\% \citep[e.g.][]{ono12, pentericci11, schenker12}.  However, it is also possible that gas within the galaxy, rather than in the IGM, also attenuates \lya.  The increasingly blue colors of galaxies in our sample could indicate that these galaxies are becoming more gas-rich towards higher redshift, as a higher gas surface density begets more star-formation, resulting in bluer colors \citep[e.g.][]{kennicutt98}.

Using observations of the star-formation rates, stellar masses and sizes of galaxies from $z =$ 2--8, \citet{papovich11} inferred the evolution of the gas accretion rate with redshift (see Figure 4 in \citet{papovich11}).  In Figure 8, we also plot their value for the ratio of the gas accretion rate (\.M$_{gas}$) to the star-formation rate (\.M$_{SFR}$).  This value is predicted to rise strongly with redshift; at $z \sim$ 3, galaxies are turning all of the accreted gas into stars, while by $z \sim$ 7, galaxies are accreting $\sim$ 2.5$\times$ more gas than they can convert into stars.  In fact, this ratio of the gas accretion rate to the star-formation rate rises by $\sim$ 40\% from $z =$ 6 to $z =$ 7.  

Therefore, one interpretation based on these model predictions and the relative blue colors of the $z\sim 7$ galaxies is that these galaxies are accreting vast reservoirs of gas.  This gas is likely to have an attenuating effect on the Ly$\alpha$ emission line.  We note that gas \textit{outflows} in galaxies have been invoked as an escape mechanism for Ly$\alpha$ emission, as Ly$\alpha$ photons escape from outflowing gas that is redshifted with respect to the systemic redshift of the galaxy \citep[e.g.,][]{verhamme08, dijkstra10}.  However, at $z \sim$ 7 if the gas outflow rate is related to the SFR, and the gas accretion rate exceeds the SFR, it stands to reason that the gas \textit{inflow} rates dominate the gas dynamics in these galaxies.  The gas inflows would have a net blueshift with respect to the galaxy systemic redshift, which would effectively block any Ly$\alpha$ emission from the galaxy.  Therefore, the decrease in the incidence of Ly$\alpha$ in galaxies from $z\sim 6$ to 7 could be explained in part by an increase in the gas accretion rate and a decrease in the gas outflow rate going to higher redshift, and thus may still be consistent with a mostly ionized IGM.

Finally, we note that this scenario can also explain the observation that the Ly$\alpha$ incidence drops more in faint galaxies than bright galaxies \citep{ono12}.  \citet{krumholz12} have shown that as the SFR depends on the metallicity, galaxies in smaller dark matter halos are less able to efficiently turn gas into stars than those in large halos.  In these smaller galaxies the suppressed SFR results in a lower metal production, which keeps the SFR much lower than the gas inflow rate, and thus likely suppresses \lya\ emission.    However, the observed mass-metallicity relation at $z =$ 0 implies that massive galaxies are more able to retain their metals, hence they would advance past the stage of not being able to keep up with inflows sooner than less-massive galaxies, which would cause their \lya\ EWs to evolve less rapidly and their metallicities (and hence dust columns) to evolve more rapidly, just as observed. 

Further statistics of the Ly$\alpha$ incidence, as well as gas mass measurements with ALMA and metallicity measurements with {\it JWST} will allow for more detailed tests of these models.

\section{Conclusions}
Using the new deep, wide-area data from the CANDELS Multi-cycle {\it HST} program, as well as existing data from the HUDF and ERS programs, we have analysed the evolution of the UV spectral slope $\beta$ with redshift.  We utilized photometric redshifts to select galaxy samples at $z =$ 4, 5, 6, 7 and 8.  We measured $\beta$ with a new technique, fitting a slope to the best-fit model spectrum during the SED fitting process in the Calzetti et al.\ (1994) defined wavelength windows.  Using simulations, we have shown that this method determines $\beta$ with smaller scatter and bias than the previously used methods at all redshifts and magnitudes.

We computed the median value of $\beta$ in each redshift bin, as well as in subsamples split by both UV luminosity and stellar mass, and reached the following conclusions:
\vspace{2mm}

$\bullet$ The median value of $\beta$ for faint galaxies at $z =$ 7 is $\beta$ = $-$2.68$^{+0.39}_{-0.24}$ ($\sim -$2.4 after correction for observational bias).  This value is redder than those previously reported in the literature ($\beta \sim$ $-$3), and has an uncertainty reduced by $\sim$ 30\%.  These combine to cast further doubt upon the existence of a large population of metal-free stars at high redshift, as $\beta$ = $-$2.7 is consistent with a young ($\lesssim$ 50 Myr), dust-free population of 0.2 $Z$\sol\ stars.
\vspace{2mm}

$\bullet$ We measure significant evolution of the median $\beta$ for galaxies at all luminosities from $z =$ 4 -- 7.  Galaxies at $z =$ 7 are again consistent with little-to-no dust, while those at $z =$ 4 have rest-UV colors consistent with significant dust extinction on the order of A$_V \sim$ 0.5 mag (depending on the exact formation redshift of the galaxies).  If the typical galaxy in our sample forms its first generation of stars at $z \sim$ 15--20, then this onset of dust extinction at $z <$ 7 is commensurate with the evolution of $M \lesssim$ 3.5 M\sol\ stars off the main sequence and through their AGB phase.  AGB stars with initial masses of $\sim$2--4 M\sol\ are the most efficient dust producers, thus this trend in $\beta$ may be indicative of the increased influence of dust formed in the atmospheres of AGB stars.  The low level of dust at $z =$ 7 might indicate, at least for the typical L $\lesssim$ L$^{\ast}$ galaxies in our sample, that dust formed in SNe plays a smaller role.
\vspace{2mm}

$\bullet$ We fit for a linear correlation between $\beta$ and M$_{1500}$ to determine whether we detect a significant $\beta$ --  UV absolute magnitude relation.  At $z \geq$ 6, we detect no significant relation, however the uncertainties on this slope are high, so a more robust determination of whether such a slope exists will need to wait for larger high-redshift samples.  However, at $z =$ 4 and 5, we also measure slopes consistent with zero at higher significance, though we note that this may be dominated by a lack of a trend at bright magnitudes, and a possible trend at fainter magnitudes.  Thus, the dynamic range in M$_{UV}$ is crucial, as when restricting our faint bin to HUDF galaxies only, we do observe a correlation.  To compare our results to those in the literature, we have run various tests, and we find that the definition of the rest-frame wavelength used for M$_{UV}$ can determine whether a correlation is observed or not.  Our method of using a constant rest-frame wavelength of 1500 \AA\ may wash out such a correlation, as we detect a significant correlation at $z =$ 4 when using 2000 \AA\ to define M$_{UV}$ (though the slope of the correlation is much less than other studies have found).  Additionally, the method of deriving $\beta$ can affect the results, as if one uses only a single color to derive $\beta$, one can observe such a $\beta$-M$_{UV}$ dependence at the 2$\sigma$ level (though we note some studies do acknowledge and correct for this).  Future work examining this correlation with a larger sample covering a wide range of M$_{UV}$, and examining all possible biases, will be crucial to understand the origin of any such correlation.

\vspace{2mm}

$\bullet$ Comparing $\beta$ to the stellar mass does result in a significant correlation, in that more massive galaxies appear redder.  This may imply that more massive galaxies experience accelerated evolution, as they appear red at all redshifts in our sample, though selection effects make this difficult to quantify at the highest redshifts.  It thus may be that SNe are efficient dust producers, but only the most massive galaxies ($>$ 10$^{9}$ M\sol) can retain this dust, while lower-mass galaxies lose the dust in outflows.  In the low mass galaxies, it is not until lower-mass stars pass through the AGB phase that their dust masses begin to build in earnest.
\vspace{2mm}

$\bullet$  We find excellent agreement between our observed evolution of $\beta$ with redshift and the same trend in the simulations of \citet{finlator11}, which replicate the observed mass-metallicity relation at lower redshifts.  These simulations assume the low-redshift value for amount of extinction per metallicity in a galaxy, thus this agreement implies that galaxies at high redshift contain a similar yield of dust-per-metal as at low redshift.  Additionally, the slope of the $\beta$ -- mass relation we measure at $z =$ 4 is consistent with that from the \citet{finlator11} simulations, implying that the observed evolution in the mass-metallicity relation continues out to at least $z =$ 4.
\vspace{2mm}

$\bullet$ The blue colors of our low-mass galaxies indicate that they should have a strong \lya\ emission line, yet recent results have shown that galaxies at $z =$ 7 have weaker than expected \lya\ emission.  Recent work indicates that galaxies are likely building vast reservoirs of gas with increasing redshift, which is plausible given the blue colors of our galaxies.  This may in turn reduce the visibility of the \lya\ line without the need to invoke a large neutral fraction in the IGM.
\vspace{2mm}
 
While the evolution of galaxy rest-frame UV color from $z =$ 4 -- 7 is robust, the mechanism behind this trend is less clear.  It is likely due to dust, but the exact interplay between dust-formation, galaxy stellar mass, and feedback requires further study.  More robust results at $z =$ 8 (i.e., utilizing data from the upcoming treasury proposal in the HUDF; PI Ellis) may shed light on the issue, allowing examination of whether the observed trends continue into that earlier epoch.

\acknowledgements
We thank Ivo Labb{\'e}, Pascal Oesch, Milos Milosavljevic and Viviana Acquaviva for helpful conversations.  We also thank the referee, Rychard Bouwens, for his thorough review which greatly improved this paper.  Support for SLF was provided by NASA through {\it HST} Cycle 18 grant HST-AR-12127 and through Hubble Fellowship grant HST-HF-51288.01.  CP was supported in part by HST programs 12060 and 12127.  These grants were all awarded by the Space Telescope Science Institute, which is operated by the Association of Universities for Research in Astronomy, Inc., for NASA, under contract NAS 5-26555.


\begin{thebibliography}{90}
\expandafter\ifx\csname natexlab\endcsname\relax\def\natexlab#1{#1}\fi

\bibitem[{{Beckwith} {et~al.}(2006){Beckwith}, {Stiavelli}, {Koekemoer},
  {Caldwell}, {Ferguson}, {Hook}, {Lucas}, {Bergeron}, {Corbin}, {Jogee},
  {Panagia}, {Robberto}, {Royle}, {Somerville}, \& {Sosey}}]{beckwith06}
{Beckwith}, S.~V.~W., {Stiavelli}, M., {Koekemoer}, A.~M., {et~al.} 2006, \aj,
  132, 1729

\bibitem[{{Bertin} \& {Arnouts}(1996)}]{bertin96}
{Bertin}, E., \& {Arnouts}, S. 1996, \aaps, 117, 393

\bibitem[{{Bouwens} {et~al.}(2006){Bouwens}, {Illingworth}, {Blakeslee}, \&
  {Franx}}]{bouwens06}
{Bouwens}, R.~J., {Illingworth}, G.~D., {Blakeslee}, J.~P., \& {Franx}, M.
  2006, \apj, 653, 53

\bibitem[{{Bouwens} {et~al.}(2007){Bouwens}, {Illingworth}, {Franx}, \&
  {Ford}}]{bouwens07}
{Bouwens}, R.~J., {Illingworth}, G.~D., {Franx}, M., \& {Ford}, H. 2007, \apj,
  670, 928

\bibitem[{{Bouwens} {et~al.}(2009){Bouwens}, {Illingworth}, {Franx}, {Chary},
  {Meurer}, {Conselice}, {Ford}, {Giavalisco}, \& {van Dokkum}}]{bouwens09}
{Bouwens}, R.~J., {Illingworth}, G.~D., {Franx}, M., {et~al.} 2009, \apj, 705,
  936

\bibitem[{{Bouwens} {et~al.}(2010{\natexlab{a}}){Bouwens}, {Illingworth},
  {Oesch}, {Stiavelli}, {van Dokkum}, {Trenti}, {Magee}, {Labb{\'e}}, {Franx},
  {Carollo}, \& {Gonzalez}}]{bouwens10a}
{Bouwens}, R.~J., {Illingworth}, G.~D., {Oesch}, P.~A., {et~al.}
  2010{\natexlab{a}}, \apjl, 709, L133

\bibitem[{{Bouwens} {et~al.}(2010{\natexlab{b}}){Bouwens}, {Illingworth},
  {Oesch}, {Trenti}, {Stiavelli}, {Carollo}, {Franx}, {van Dokkum},
  {Labb{\'e}}, \& {Magee}}]{bouwens10b}
---. 2010{\natexlab{b}}, \apjl, 708, L69

\bibitem[{{Bouwens} {et~al.}(2011){Bouwens}, {Illingworth}, {Oesch},
  {Labb{\'e}}, {Trenti}, {van Dokkum}, {Franx}, {Stiavelli}, {Carollo},
  {Magee}, \& {Gonzalez}}]{bouwens11}
---. 2011, \apj, 737, 90

\bibitem[{{Bouwens} {et~al.}(2012){Bouwens}, {Illingworth}, {Oesch}, {Franx},
  {Labbe}, {Trenti}, {van Dokkum}, {Carollo}, {Gonzalez}, \&
  {Magee}}]{bouwens12}
---. 2012, ArXiv e-prints, astroph/1109.0994

\bibitem[{{Brammer} {et~al.}(2008){Brammer}, {van Dokkum}, \&
  {Coppi}}]{brammer08}
{Brammer}, G.~B., {van Dokkum}, P.~G., \& {Coppi}, P. 2008, \apj, 686, 1503

\bibitem[{{Bruzual} \& {Charlot}(2003)}]{bruzual03}
{Bruzual}, G., \& {Charlot}, S. 2003, \mnras, 344, 1000

\bibitem[{{Bunker} {et~al.}(2010){Bunker}, {Wilkins}, {Ellis}, {Stark},
  {Lorenzoni}, {Chiu}, {Lacy}, {Jarvis}, \& {Hickey}}]{bunker10}
{Bunker}, A.~J., {Wilkins}, S., {Ellis}, R.~S., {et~al.} 2010, \mnras, 409, 855

\bibitem[{{Calzetti} {et~al.}(2000){Calzetti}, {Armus}, {Bohlin}, {Kinney},
  {Koornneef}, \& {Storchi-Bergmann}}]{calzetti00}
{Calzetti}, D., {Armus}, L., {Bohlin}, R.~C., {et~al.} 2000, \apj, 533, 682

\bibitem[{{Calzetti} {et~al.}(1994){Calzetti}, {Kinney}, \&
  {Storchi-Bergmann}}]{calzetti94}
{Calzetti}, D., {Kinney}, A.~L., \& {Storchi-Bergmann}, T. 1994, \apj, 429, 582

\bibitem[{{Cardelli} {et~al.}(1989){Cardelli}, {Clayton}, \&
  {Mathis}}]{cardelli89}
{Cardelli}, J.~A., {Clayton}, G.~C., \& {Mathis}, J.~S. 1989, \apj, 345, 245

\bibitem[{{Castellano} {et~al.}(2012){Castellano}, {Fontana}, {Grazian},
  {Pentericci}, {Santini}, {Koekemoer}, {Cristiani}, {Galametz}, {Gallerani},
  {Vanzella}, {Boutsia}, {Gallozzi}, {Giallongo}, {Maiolino}, {Menci}, \&
  {Paris}}]{castellano12}
{Castellano}, M., {Fontana}, A., {Grazian}, A., {et~al.} 2012, \aap, 540, A39

\bibitem[{{Dav{\'e}} {et~al.}(2006){Dav{\'e}}, {Finlator}, \&
  {Oppenheimer}}]{dave06}
{Dav{\'e}}, R., {Finlator}, K., \& {Oppenheimer}, B.~D. 2006, \mnras, 370, 273

\bibitem[{{Dav{\'e}} {et~al.}(2011){Dav{\'e}}, {Finlator}, \&
  {Oppenheimer}}]{dave11}
---. 2011, \mnras, 416, 1354

\bibitem[{{Dickinson} {et~al.}(2003){Dickinson}, {Giavalisco}, \& {The Goods
  Team}}]{dickinson03}
{Dickinson}, M., {Giavalisco}, M., \& {The Goods Team}. 2003, in The Mass of
  Galaxies at Low and High Redshift, ed. {R.~Bender \& A.~Renzini}, 324

\bibitem[{{Dijkstra} \& {Wyithe}(2010)}]{dijkstra10}
{Dijkstra}, M., \& {Wyithe}, J.~S.~B. 2010, \mnras, 408, 352

\bibitem[{{Draine}(2009)}]{draine09}
{Draine}, B.~T. 2009, in Astronomical Society of the Pacific Conference Series,
  Vol. 414, Cosmic Dust - Near and Far, ed. {T.~Henning, E.~Gr{\"u}n, \&
  J.~Steinacker}, 453

\bibitem[{{Draine} \& {Salpeter}(1979)}]{draine79}
{Draine}, B.~T., \& {Salpeter}, E.~E. 1979, \apj, 231, 438

\bibitem[{{Dunlop} {et~al.}(2012){Dunlop}, {McLure}, {Robertson}, {Ellis},
  {Stark}, {Cirasuolo}, \& {de Ravel}}]{dunlop12}
{Dunlop}, J.~S., {McLure}, R.~J., {Robertson}, B.~E., {et~al.} 2012, \mnras,
  420, 901

\bibitem[{{Dwek} \& {Scalo}(1980)}]{dwek80}
{Dwek}, E., \& {Scalo}, J.~M. 1980, \apj, 239, 193

\bibitem[{{Erb} {et~al.}(2010){Erb}, {Pettini}, {Shapley}, {Steidel}, {Law}, \&
  {Reddy}}]{erb10}
{Erb}, D.~K., {Pettini}, M., {Shapley}, A.~E., {et~al.} 2010, \apj, 719, 1168

\bibitem[{{Erb} {et~al.}(2006){Erb}, {Steidel}, {Shapley}, {Pettini}, {Reddy},
  \& {Adelberger}}]{erb06}
{Erb}, D.~K., {Steidel}, C.~C., {Shapley}, A.~E., {et~al.} 2006, \apj, 646, 107

\bibitem[{{Fazio} {et~al.}(2004){Fazio}, {Hora}, {Allen}, {Ashby}, {Barmby},
  {Deutsch}, {Huang}, {Kleiner}, {Marengo}, {Megeath}, {Melnick}, {Pahre},
  {Patten}, {Polizotti}, {Smith}, {Taylor}, {Wang}, {Willner}, {Hoffmann},
  {Pipher}, {Forrest}, {McMurty}, {McCreight}, {McKelvey}, {McMurray}, {Koch},
  {Moseley}, {Arendt}, {Mentzell}, {Marx}, {Losch}, {Mayman}, {Eichhorn},
  {Krebs}, {Jhabvala}, {Gezari}, {Fixsen}, {Flores}, {Shakoorzadeh}, {Jungo},
  {Hakun}, {Workman}, {Karpati}, {Kichak}, {Whitley}, {Mann}, {Tollestrup},
  {Eisenhardt}, {Stern}, {Gorjian}, {Bhattacharya}, {Carey}, {Nelson},
  {Glaccum}, {Lacy}, {Lowrance}, {Laine}, {Reach}, {Stauffer}, {Surace},
  {Wilson}, {Wright}, {Hoffman}, {Domingo}, \& {Cohen}}]{fazio04}
{Fazio}, G.~G., {Hora}, J.~L., {Allen}, L.~E., {et~al.} 2004, \apjs, 154, 10

\bibitem[{{Finkelstein} {et~al.}(2010){Finkelstein}, {Papovich}, {Giavalisco},
  {Reddy}, {Ferguson}, {Koekemoer}, \& {Dickinson}}]{finkelstein10a}
{Finkelstein}, S.~L., {Papovich}, C., {Giavalisco}, M., {et~al.} 2010, \apj,
  719, 1250

\bibitem[{{Finkelstein} {et~al.}(2009){Finkelstein}, {Rhoads}, {Malhotra}, \&
  {Grogin}}]{finkelstein09a}
{Finkelstein}, S.~L., {Rhoads}, J.~E., {Malhotra}, S., \& {Grogin}, N. 2009,
  \apj, 691, 465

\bibitem[{{Finlator} \& {Dav{\'e}}(2008)}]{finlator08}
{Finlator}, K., \& {Dav{\'e}}, R. 2008, \mnras, 385, 2181

\bibitem[{{Finlator} {et~al.}(2006){Finlator}, {Dav{\'e}}, {Papovich}, \&
  {Hernquist}}]{finlator06}
{Finlator}, K., {Dav{\'e}}, R., {Papovich}, C., \& {Hernquist}, L. 2006, \apj,
  639, 672

\bibitem[{{Finlator} {et~al.}(2011){Finlator}, {Oppenheimer}, \&
  {Dav{\'e}}}]{finlator11}
{Finlator}, K., {Oppenheimer}, B.~D., \& {Dav{\'e}}, R. 2011, \mnras, 410, 1703

\bibitem[{{Fioc} \& {Rocca-Volmerange}(1997)}]{fioc97}
{Fioc}, M., \& {Rocca-Volmerange}, B. 1997, \aap, 326, 950

\bibitem[{{Gall} {et~al.}(2011){Gall}, {Hjorth}, \& {Andersen}}]{gall11}
{Gall}, C., {Hjorth}, J., \& {Andersen}, A.~C. 2011, \aapr, 19, 43

\bibitem[{{Gehrz}(1989)}]{gehrz89}
{Gehrz}, R. 1989, in IAU Symposium, Vol. 135, Interstellar Dust, ed.
  {L.~J.~Allamandola \& A.~G.~G.~M.~Tielens}, 445

\bibitem[{{Giavalisco} {et~al.}(2004){Giavalisco}, {Dickinson}, {Ferguson},
  {Ravindranath}, {Kretchmer}, {Moustakas}, {Madau}, {Fall}, {Gardner},
  {Livio}, {Papovich}, {Renzini}, {Spinrad}, {Stern}, \&
  {Riess}}]{giavalisco04}
{Giavalisco}, M., {Dickinson}, M., {Ferguson}, H.~C., {et~al.} 2004, \apjl,
  600, L103

\bibitem[{{Grazian} {et~al.}(2006){Grazian}, {Fontana}, {de Santis}, {Nonino},
  {Salimbeni}, {Giallongo}, {Cristiani}, {Gallozzi}, \& {Vanzella}}]{grazian06}
{Grazian}, A., {Fontana}, A., {de Santis}, C., {et~al.} 2006, \aap, 449, 951

\bibitem[{{Grogin} {et~al.}(2011){Grogin}, {Kocevski}, {Faber}, {Ferguson},
  {Koekemoer}, {Riess}, {Acquaviva}, {Alexander}, {Almaini}, {Ashby}, {Barden},
  {Bell}, {Bournaud}, {Brown}, {Caputi}, {Casertano}, {Cassata}, {Castellano},
  {Challis}, {Chary}, {Cheung}, {Cirasuolo}, {Conselice}, {Roshan Cooray},
  {Croton}, {Daddi}, {Dahlen}, {Dav{\'e}}, {de Mello}, {Dekel}, {Dickinson},
  {Dolch}, {Donley}, {Dunlop}, {Dutton}, {Elbaz}, {Fazio}, {Filippenko},
  {Finkelstein}, {Fontana}, {Gardner}, {Garnavich}, {Gawiser}, {Giavalisco},
  {Grazian}, {Guo}, {Hathi}, {H{\"a}ussler}, {Hopkins}, {Huang}, {Huang},
  {Jha}, {Kartaltepe}, {Kirshner}, {Koo}, {Lai}, {Lee}, {Li}, {Lotz}, {Lucas},
  {Madau}, {McCarthy}, {McGrath}, {McIntosh}, {McLure}, {Mobasher},
  {Moustakas}, {Mozena}, {Nandra}, {Newman}, {Niemi}, {Noeske}, {Papovich},
  {Pentericci}, {Pope}, {Primack}, {Rajan}, {Ravindranath}, {Reddy}, {Renzini},
  {Rix}, {Robaina}, {Rodney}, {Rosario}, {Rosati}, {Salimbeni}, {Scarlata},
  {Siana}, {Simard}, {Smidt}, {Somerville}, {Spinrad}, {Straughn}, {Strolger},
  {Telford}, {Teplitz}, {Trump}, {van der Wel}, {Villforth}, {Wechsler},
  {Weiner}, {Wiklind}, {Wild}, {Wilson}, {Wuyts}, {Yan}, \& {Yun}}]{grogin11}
{Grogin}, N.~A., {Kocevski}, D.~D., {Faber}, S.~M., {et~al.} 2011, \apjs, 197,
  35

\bibitem[{{Hathi} {et~al.}(2008){Hathi}, {Malhotra}, \& {Rhoads}}]{hathi08}
{Hathi}, N.~P., {Malhotra}, S., \& {Rhoads}, J.~E. 2008, \apj, 673, 686

\bibitem[{{Inoue}(2011)}]{inoue11}
{Inoue}, A.~K. 2011, \mnras, 415, 2920

\bibitem[{{Kalirai} {et~al.}(2009){Kalirai}, {MacKenty}, {Bohlin}, {Brown},
  {Deustua}, {Kimble}, \& {Riess}}]{kalirai09}
{Kalirai}, J.~S., {MacKenty}, J., {Bohlin}, R., {et~al.} 2009, {WFC3 SMOV
  Proposal 11451: The Photometric Performance and Calibration of WFC3/IR},
  Tech. rep.

\bibitem[{{Kennicutt}(1998)}]{kennicutt98}
{Kennicutt}, Jr., R.~C. 1998, \araa, 36, 189

\bibitem[{{Kinney} {et~al.}(1996){Kinney}, {Calzetti}, {Bohlin}, {McQuade},
  {Storchi-Bergmann}, \& {Schmitt}}]{kinney96}
{Kinney}, A.~L., {Calzetti}, D., {Bohlin}, R.~C., {et~al.} 1996, \apj, 467, 38

\bibitem[{{Koekemoer} {et~al.}(2011){Koekemoer}, {Faber}, {Ferguson}, {Grogin},
  {Kocevski}, {Koo}, {Lai}, {Lotz}, {Lucas}, {McGrath}, {Ogaz}, {Rajan},
  {Riess}, {Rodney}, {Strolger}, {Casertano}, {Castellano}, {Dahlen},
  {Dickinson}, {Dolch}, {Fontana}, {Giavalisco}, {Grazian}, {Guo}, {Hathi},
  {Huang}, {van der Wel}, {Yan}, {Acquaviva}, {Alexander}, {Almaini}, {Ashby},
  {Barden}, {Bell}, {Bournaud}, {Brown}, {Caputi}, {Cassata}, {Challis},
  {Chary}, {Cheung}, {Cirasuolo}, {Conselice}, {Roshan Cooray}, {Croton},
  {Daddi}, {Dav{\'e}}, {de Mello}, {de Ravel}, {Dekel}, {Donley}, {Dunlop},
  {Dutton}, {Elbaz}, {Fazio}, {Filippenko}, {Finkelstein}, {Frazer}, {Gardner},
  {Garnavich}, {Gawiser}, {Gruetzbauch}, {Hartley}, {H{\"a}ussler},
  {Herrington}, {Hopkins}, {Huang}, {Jha}, {Johnson}, {Kartaltepe},
  {Khostovan}, {Kirshner}, {Lani}, {Lee}, {Li}, {Madau}, {McCarthy},
  {McIntosh}, {McLure}, {McPartland}, {Mobasher}, {Moreira}, {Mortlock},
  {Moustakas}, {Mozena}, {Nandra}, {Newman}, {Nielsen}, {Niemi}, {Noeske},
  {Papovich}, {Pentericci}, {Pope}, {Primack}, {Ravindranath}, {Reddy},
  {Renzini}, {Rix}, {Robaina}, {Rosario}, {Rosati}, {Salimbeni}, {Scarlata},
  {Siana}, {Simard}, {Smidt}, {Snyder}, {Somerville}, {Spinrad}, {Straughn},
  {Telford}, {Teplitz}, {Trump}, {Vargas}, {Villforth}, {Wagner}, {Wandro},
  {Wechsler}, {Weiner}, {Wiklind}, {Wild}, {Wilson}, {Wuyts}, \&
  {Yun}}]{koekemoer11}
{Koekemoer}, A.~M., {Faber}, S.~M., {Ferguson}, H.~C., {et~al.} 2011, \apjs,
  197, 36

\bibitem[{{Krumholz} \& {Dekel}(2012)}]{krumholz12}
{Krumholz}, M.~R., \& {Dekel}, A. 2012, \apj, 753, 16

\bibitem[{{Labb{\'e}} {et~al.}(2007){Labb{\'e}}, {Franx}, {Rudnick},
  {Schreiber}, {van Dokkum}, {Moorwood}, {Rix}, {R{\"o}ttgering}, {Trujillo},
  \& {van der Werf}}]{labbe03}
{Labb{\'e}}, I., {Franx}, M., {Rudnick}, G., {et~al.} 2007, \apj, 665, 944

\bibitem[{{Labb{\'e}} {et~al.}(2010{\natexlab{a}}){Labb{\'e}}, {Gonz{\'a}lez},
  {Bouwens}, {Illingworth}, {Franx}, {Trenti}, {Oesch}, {van Dokkum},
  {Stiavelli}, {Carollo}, {Kriek}, \& {Magee}}]{labbe10b}
{Labb{\'e}}, I., {Gonz{\'a}lez}, V., {Bouwens}, R.~J., {et~al.}
  2010{\natexlab{a}}, \apjl, 716, L103

\bibitem[{{Labb{\'e}} {et~al.}(2010{\natexlab{b}}){Labb{\'e}}, {Gonz{\'a}lez},
  {Bouwens}, {Illingworth}, {Oesch}, {van Dokkum}, {Carollo}, {Franx},
  {Stiavelli}, {Trenti}, {Magee}, \& {Kriek}}]{labbe10}
---. 2010{\natexlab{b}}, \apjl, 708, L26

\bibitem[{{Laidler} {et~al.}(2007){Laidler}, {Papovich}, {Grogin}, {Idzi},
  {Dickinson}, {Ferguson}, {Hilbert}, {Clubb}, \& {Ravindranath}}]{laidler07}
{Laidler}, V.~G., {Papovich}, C., {Grogin}, N.~A., {et~al.} 2007, \pasp, 119,
  1325

\bibitem[{{Lee} {et~al.}(2011){Lee}, {Dey}, {Reddy}, {Brown}, {Gonzalez},
  {Jannuzi}, {Cooper}, {Fan}, {Bian}, {Glikman}, {Stern}, {Brodwin}, \&
  {Cooray}}]{lee11}
{Lee}, K.-S., {Dey}, A., {Reddy}, N., {et~al.} 2011, \apj, 733, 99

\bibitem[{{Lehnert} \& {Bremer}(2003)}]{lehnert03}
{Lehnert}, M.~D., \& {Bremer}, M. 2003, \apj, 593, 630

\bibitem[{{Madau}(1995)}]{madau95}
{Madau}, P. 1995, \apj, 441, 18

\bibitem[{{McLure} {et~al.}(2010){McLure}, {Dunlop}, {Cirasuolo}, {Koekemoer},
  {Sabbi}, {Stark}, {Targett}, \& {Ellis}}]{mclure10}
{McLure}, R.~J., {Dunlop}, J.~S., {Cirasuolo}, M., {et~al.} 2010, \mnras, 403,
  960

\bibitem[{{McLure} {et~al.}(2011){McLure}, {Dunlop}, {de Ravel}, {Cirasuolo},
  {Ellis}, {Schenker}, {Robertson}, {Koekemoer}, {Stark}, \&
  {Bowler}}]{mclure11}
{McLure}, R.~J., {Dunlop}, J.~S., {de Ravel}, L., {et~al.} 2011, \mnras, 418,
  2074

\bibitem[{{Meurer} {et~al.}(1999){Meurer}, {Heckman}, \& {Calzetti}}]{meurer99}
{Meurer}, G.~R., {Heckman}, T.~M., \& {Calzetti}, D. 1999, \apj, 521, 64

\bibitem[{{Meurer} {et~al.}(1997){Meurer}, {Heckman}, {Lehnert}, {Leitherer},
  \& {Lowenthal}}]{meurer97}
{Meurer}, G.~R., {Heckman}, T.~M., {Lehnert}, M.~D., {Leitherer}, C., \&
  {Lowenthal}, J. 1997, \aj, 114, 54

\bibitem[{{Micha{\l}owski} {et~al.}(2010){Micha{\l}owski}, {Murphy}, {Hjorth},
  {Watson}, {Gall}, \& {Dunlop}}]{michalowski10}
{Micha{\l}owski}, M.~J., {Murphy}, E.~J., {Hjorth}, J., {et~al.} 2010, \aap,
  522, A15

\bibitem[{{Oesch} {et~al.}(2007){Oesch}, {Stiavelli}, {Carollo}, {Bergeron},
  {Koekemoer}, {Lucas}, {Pavlovsky}, {Trenti}, {Lilly}, {Beckwith}, {Dahlen},
  {Ferguson}, {Gardner}, {Lacey}, {Mobasher}, {Panagia}, \& {Rix}}]{oesch07}
{Oesch}, P.~A., {Stiavelli}, M., {Carollo}, C.~M., {et~al.} 2007, \apj, 671,
  1212

\bibitem[{{Oesch} {et~al.}(2010){Oesch}, {Bouwens}, {Illingworth}, {Carollo},
  {Franx}, {Labb{\'e}}, {Magee}, {Stiavelli}, {Trenti}, \& {van
  Dokkum}}]{oesch10}
{Oesch}, P.~A., {Bouwens}, R.~J., {Illingworth}, G.~D., {et~al.} 2010, \apjl,
  709, L16

\bibitem[{{Ono} {et~al.}(2012){Ono}, {Ouchi}, {Mobasher}, {Dickinson},
  {Penner}, {Shimasaku}, {Weiner}, {Kartaltepe}, {Nakajima}, {Nayyeri},
  {Stern}, {Kashikawa}, \& {Spinrad}}]{ono12}
{Ono}, Y., {Ouchi}, M., {Mobasher}, B., {et~al.} 2012, \apj, 744, 83

\bibitem[{{Osterbrock} \& {Ferland}(2006)}]{osterbrock06}
{Osterbrock}, D.~E., \& {Ferland}, G.~J. 2006, {Astrophysics of gaseous nebulae
  and active galactic nuclei}, ed. {Osterbrock, D.~E.~\& Ferland, G.~J.}

\bibitem[{{Overzier} {et~al.}(2008){Overzier}, {Bouwens}, {Cross}, {Venemans},
  {Miley}, {Zirm}, {Ben{\'{\i}}tez}, {Blakeslee}, {Coe}, {Demarco}, {Ford},
  {Homeier}, {Illingworth}, {Kurk}, {Martel}, {Mei}, {Oliveira},
  {R{\"o}ttgering}, {Tsvetanov}, \& {Zheng}}]{overzier08}
{Overzier}, R.~A., {Bouwens}, R.~J., {Cross}, N.~J.~G., {et~al.} 2008, \apj,
  673, 143

\bibitem[{{Papovich} {et~al.}(2001){Papovich}, {Dickinson}, \&
  {Ferguson}}]{papovich01}
{Papovich}, C., {Dickinson}, M., \& {Ferguson}, H.~C. 2001, \apj, 559, 620

\bibitem[{{Papovich} {et~al.}(2011){Papovich}, {Finkelstein}, {Ferguson},
  {Lotz}, \& {Giavalisco}}]{papovich11}
{Papovich}, C., {Finkelstein}, S.~L., {Ferguson}, H.~C., {Lotz}, J.~M., \&
  {Giavalisco}, M. 2011, \mnras, 412, 1123

\bibitem[{{Papovich} {et~al.}(2004){Papovich}, {Dickinson}, {Ferguson},
  {Giavalisco}, {Lotz}, {Madau}, {Idzi}, {Kretchmer}, {Moustakas}, {de Mello},
  {Gardner}, {Rieke}, {Somerville}, \& {Stern}}]{papovich04}
{Papovich}, C., {Dickinson}, M., {Ferguson}, H.~C., {et~al.} 2004, \apjl, 600,
  L111

\bibitem[{{Peng} {et~al.}(2002){Peng}, {Ho}, {Impey}, \& {Rix}}]{peng02}
{Peng}, C.~Y., {Ho}, L.~C., {Impey}, C.~D., \& {Rix}, H.-W. 2002, \aj, 124, 266

\bibitem[{{Pentericci} {et~al.}(2011){Pentericci}, {Fontana}, {Vanzella},
  {Castellano}, {Grazian}, {Dijkstra}, {Boutsia}, {Cristiani}, {Dickinson},
  {Giallongo}, {Giavalisco}, {Maiolino}, {Moorwood}, {Paris}, \&
  {Santini}}]{pentericci11}
{Pentericci}, L., {Fontana}, A., {Vanzella}, E., {et~al.} 2011, \apj, 743, 132

\bibitem[{{Reddy} {et~al.}(2012){Reddy}, {Dickinson}, {Elbaz}, {Morrison},
  {Giavalisco}, {Ivison}, {Papovich}, {Scott}, {Buat}, {Burgarella},
  {Charmandaris}, {Daddi}, {Magdis}, {Murphy}, {Altieri}, {Aussel},
  {Dannerbauer}, {Dasyra}, {Hwang}, {Kartaltepe}, {Leiton}, {Magnelli}, \&
  {Popesso}}]{reddy12}
{Reddy}, N., {Dickinson}, M., {Elbaz}, D., {et~al.} 2012, \apj, 744, 154

\bibitem[{{Reddy} {et~al.}(2006){Reddy}, {Steidel}, {Erb}, {Shapley}, \&
  {Pettini}}]{reddy06}
{Reddy}, N.~A., {Steidel}, C.~C., {Erb}, D.~K., {Shapley}, A.~E., \& {Pettini},
  M. 2006, \apj, 653, 1004

\bibitem[{{Reddy} {et~al.}(2008){Reddy}, {Steidel}, {Pettini}, {Adelberger},
  {Shapley}, {Erb}, \& {Dickinson}}]{reddy08}
{Reddy}, N.~A., {Steidel}, C.~C., {Pettini}, M., {et~al.} 2008, \apjs, 175, 48

\bibitem[{{Schaerer}(2002)}]{schaerer02}
{Schaerer}, D. 2002, \aap, 382, 28

\bibitem[{{Schaerer} \& {de Barros}(2010)}]{schaerer10}
{Schaerer}, D., \& {de Barros}, S. 2010, \aap, 515, A73

\bibitem[{{Schenker} {et~al.}(2012){Schenker}, {Stark}, {Ellis}, {Robertson},
  {Dunlop}, {McLure}, {Kneib}, \& {Richard}}]{schenker12}
{Schenker}, M.~A., {Stark}, D.~P., {Ellis}, R.~S., {et~al.} 2012, \apj, 744,
  179

\bibitem[{{Seibert} {et~al.}(2002){Seibert}, {Heckman}, \&
  {Meurer}}]{seibert02}
{Seibert}, M., {Heckman}, T.~M., \& {Meurer}, G.~R. 2002, \aj, 124, 46

\bibitem[{{Shim} {et~al.}(2011){Shim}, {Chary}, {Dickinson}, {Lin}, {Spinrad},
  {Stern}, \& {Yan}}]{shim11}
{Shim}, H., {Chary}, R.-R., {Dickinson}, M., {et~al.} 2011, \apj, 738, 69

\bibitem[{{Stanway} {et~al.}(2005){Stanway}, {McMahon}, \&
  {Bunker}}]{stanway05}
{Stanway}, E.~R., {McMahon}, R.~G., \& {Bunker}, A.~J. 2005, \mnras, 359, 1184

\bibitem[{{Stark} {et~al.}(2009){Stark}, {Ellis}, {Bunker}, {Bundy}, {Targett},
  {Benson}, \& {Lacy}}]{stark09}
{Stark}, D.~P., {Ellis}, R.~S., {Bunker}, A., {et~al.} 2009, \apj, 697, 1493

\bibitem[{{Stark} {et~al.}(2010){Stark}, {Ellis}, {Chiu}, {Ouchi}, \&
  {Bunker}}]{stark10}
{Stark}, D.~P., {Ellis}, R.~S., {Chiu}, K., {Ouchi}, M., \& {Bunker}, A. 2010,
  \mnras, 408, 1628

\bibitem[{{Stark} {et~al.}(2011){Stark}, {Ellis}, \& {Ouchi}}]{stark11}
{Stark}, D.~P., {Ellis}, R.~S., \& {Ouchi}, M. 2011, \apjl, 728, L2

\bibitem[{{Todini} \& {Ferrara}(2001)}]{todini01}
{Todini}, P., \& {Ferrara}, A. 2001, \mnras, 325, 726

\bibitem[{{Tremonti} {et~al.}(2004){Tremonti}, {Heckman}, {Kauffmann},
  {Brinchmann}, {Charlot}, {White}, {Seibert}, {Peng}, {Schlegel}, {Uomoto},
  {Fukugita}, \& {Brinkmann}}]{tremonti04}
{Tremonti}, C.~A., {Heckman}, T.~M., {Kauffmann}, G., {et~al.} 2004, \apj, 613,
  898

\bibitem[{{Vanzella} {et~al.}(2008){Vanzella}, {Cristiani}, {Dickinson},
  {Giavalisco}, {Kuntschner}, {Haase}, {Nonino}, {Rosati}, {Cesarsky},
  {Ferguson}, {Fosbury}, {Grazian}, {Moustakas}, {Rettura}, {Popesso},
  {Renzini}, {Stern}, \& {GOODS Team}}]{vanzella08}
{Vanzella}, E., {Cristiani}, S., {Dickinson}, M., {et~al.} 2008, \aap, 478, 83

\bibitem[{{Varosi} \& {Landsman}(1993)}]{varosi93}
{Varosi}, F., \& {Landsman}, W.~B. 1993, in Astronomical Society of the Pacific
  Conference Series, Vol.~52, Astronomical Data Analysis Software and Systems
  II, ed. {R.~J.~Hanisch, R.~J.~V.~Brissenden, \& J.~Barnes}, 515

\bibitem[{{Verhamme} {et~al.}(2008){Verhamme}, {Schaerer}, {Atek}, \&
  {Tapken}}]{verhamme08}
{Verhamme}, A., {Schaerer}, D., {Atek}, H., \& {Tapken}, C. 2008, \aap, 491, 89

\bibitem[{{Wang} {et~al.}(2008){Wang}, {Carilli}, {Wagg}, {Bertoldi}, {Walter},
  {Menten}, {Omont}, {Cox}, {Strauss}, {Fan}, {Jiang}, \& {Schneider}}]{wang08}
{Wang}, R., {Carilli}, C.~L., {Wagg}, J., {et~al.} 2008, \apj, 687, 848

\bibitem[{{Wilkins} {et~al.}(2011){Wilkins}, {Bunker}, {Stanway}, {Lorenzoni},
  \& {Caruana}}]{wilkins11}
{Wilkins}, S.~M., {Bunker}, A.~J., {Stanway}, E., {Lorenzoni}, S., \&
  {Caruana}, J. 2011, \mnras, 417, 717

\bibitem[{{Windhorst} {et~al.}(2011){Windhorst}, {Cohen}, {Hathi}, {McCarthy},
  {Ryan}, {Yan}, {Baldry}, {Driver}, {Frogel}, {Hill}, {Kelvin}, {Koekemoer},
  {Mechtley}, {O'Connell}, {Robotham}, {Rutkowski}, {Seibert}, {Straughn},
  {Tuffs}, {Balick}, {Bond}, {Bushouse}, {Calzetti}, {Crockett}, {Disney},
  {Dopita}, {Hall}, {Holtzman}, {Kaviraj}, {Kimble}, {MacKenty}, {Mutchler},
  {Paresce}, {Saha}, {Silk}, {Trauger}, {Walker}, {Whitmore}, \&
  {Young}}]{windhorst11}
{Windhorst}, R.~A., {Cohen}, S.~H., {Hathi}, N.~P., {et~al.} 2011, \apjs, 193,
  27

\bibitem[{{Wuyts} {et~al.}(2008){Wuyts}, {Labb{\'e}}, {Schreiber}, {Franx},
  {Rudnick}, {Brammer}, \& {van Dokkum}}]{wuyts08}
{Wuyts}, S., {Labb{\'e}}, I., {Schreiber}, N.~M.~F., {et~al.} 2008, \apj, 682,
  985

\bibitem[{{Yajima} {et~al.}(2011){Yajima}, {Choi}, \& {Nagamine}}]{yajima11}
{Yajima}, H., {Choi}, J.-H., \& {Nagamine}, K. 2011, \mnras, 412, 411

\bibitem[{{Yan} {et~al.}(2010){Yan}, {Windhorst}, {Hathi}, {Cohen}, {Ryan},
  {O'Connell}, \& {McCarthy}}]{yan10}
{Yan}, H.-J., {Windhorst}, R.~A., {Hathi}, N.~P., {et~al.} 2010, Research in
  Astronomy and Astrophysics, 10, 867

\end{thebibliography}
\end{document}